\documentclass{aa}  

\usepackage{graphicx}
\usepackage[table,xcdraw]{xcolor}
\usepackage{natbib}
\usepackage{hyperref}
\usepackage{comment}
\usepackage{capt-of}
\usepackage{tabularx}
\usepackage{array}
\usepackage{soul}
\bibpunct{(}{)}{;}{a}{}{,} 
\newcommand{\mycomment}[1]{}

\renewcommand{\textcolor}[2]{{\textbf{#2}}}
\usepackage[T1]{fontenc}
\usepackage{lmodern}
\begin{document}

   \title{1-arcsecond imaging of ELAIS-N1 field at 144MHz using the LoTSS survey with international LOFAR telescope}

   \author{Haoyang Ye
          \inst{1,2},
          Frits Sweijen
          \inst{3,4},
          Reinout J. van Weeren
          \inst{1},
          Wendy Williams
          \inst{5},
          Jurjen de Jong
          \inst{1},
          Leah K. Morabito
          \inst{3,4}, 
          Huub Rottgering
          \inst{1},
          Timothy. W. Shimwell
          \inst{1,6},
          P.N. Best
          \inst{7},
          Marco Bondi
          \inst{8},
          Marcus Br{\"u}ggen
          \inst{9},
          Francesco de Gasperin
          \inst{8},
          \and
          Cyril Tasse
          \inst{10}
          }

   \institute{Leiden Observatory, Leiden University, PO Box 9513, 2300 RA Leiden, The Netherlands\\ \email{} hye@strw.leidenuniv.nl
   \and University of Cambridge, Cavendish Asterophysics group, JJ Thomson Avenue, Cambridge, CB3 0HE, UK\\ \email{} hy297@cam.ac.uk
   \and Centre for Extragalactic Astronomy, Department of Physics, Durham University, Department of Physics, South Road, Durham DH1 3LE, UK
   \and Institute for Computational Cosmology, Department of Physics, Durham University, South Road, Durham DH1 3LE, UK
   \and SKA Observatory, Jordrell Bank, Lower Withington, Macclesfield, SK11 9FT, UK
   \and ASTRON, the Netherlands Institute for Radio Astronomy, Oude Hoogeveensedijk 4, 7991 PD Dwingeloo, The Netherlands
   \and Institute for Astronomy, University of Edinburgh, Royal Observatory, Blackford Hill, Edinburgh, EH9 3HJ, UK
   \and INAF - Istituto di Radioastronomia, via P. Gobetti 101, 40129, Bologna, Italy
   \and Hamburg Observatory, University of Hamburg, Gojenbergsweg 112, 21029 Hamburg
   \and GEPI \& ORN, Observatoire de Paris, Université PSL, CNRS, 5 Place Jules Janssen, 92190 Meudon, France and Department of Physics \& Electronics, Rhodes University, PO Box 94, Grahamstown, 6140, South Africa
   }

   \date{Resubmitted September 27, 2023}

 
  \abstract
 {We present the first wide area ($2.5 \times 2.5$ $\mathrm{deg}^2$) LOFAR High Band Antenna image at a resolution of $1.2\arcsec\times2\arcsec$ with a median noise of $\approx 80 \mu \mathrm{Jy} {\mathrm{ beam}}^{-1}$. It was made from an 8-hour International LOFAR Telescope (ILT) observation of the ELAIS-N1 field at frequencies ranging from 120 to 168\,MHz with the most up-to-date ILT imaging methods. This intermediate resolution falls between the highest possible resolution (0.3\arcsec) achievable by using all International LOFAR Telescope (ILT) baselines and the standard 6-arcsecond resolution in the LoTSS (LOFAR Two-meter Sky Survey) image products utilising the LOFAR Dutch baselines only. This is the first demonstration of the feasibility of imaging using the ILT at a resolution of $\sim$1\arcsec, which provides unique information on source morphology at scales that fall below the surface brightness limits at higher resolutions. The total calibration and imaging computational time is approximately 52,000 core hours, nearly 5 times more than required to produce a $6\arcsec$ resolution image. 
 We also present a radio source catalogue containing 2263 sources detected over the $2.5 \times 2.5$ $\mathrm{deg}^2$ image of the ELAIS-N1 field, with a peak intensity threshold of $5.5\sigma$. The catalogue has been cross-matched with the LoTSS deep ELAIS-N1 field radio catalogue, and its flux density and positional accuracy have been investigated and corrected accordingly. We find that $\sim$80\% of sources which we expect to be detectable based on their peak brightness in the LoTSS $6\arcsec$ resolution image are detected in this image, which is approximately a factor of two higher than for $0.3\arcsec$ resolution imaging in the Lockman Hole, implying there is a wealth of information on these intermediate scales. }

    \titlerunning{1 arcsecond imaging strategy for the International LOFAR Telescope}
    \authorrunning{H.Ye et al.}
   \maketitle
%
\section{Introduction}

The LOw Frequency ARray (LOFAR; \citet{2013A&A...556A...2V}) is a low-frequency radio interferometer operating below 250\,MHz, with Low Band Antennas (LBAs) and High Band Antennas (HBAs) optimised for 10-80 MHz and 120-240\,MHz respectively. Utilising the HBAs, the ongoing LOFAR Two-metre Sky Survey (LoTSS; \citet{2017A&A...598A.104S}) has published two data releases \citep{2019A&A...622A...1S, 2022A&A...659A...1S} with image products at a resolution of $6\arcsec$, which corresponds to the full resolution capability of the Dutch LOFAR. With long intra-continental baselines of up to 2000\,km, the International LOFAR Telescope (ILT) gives the potential to survey wide fields at an angular resolution of a few tenths of an arcsecond using the HBAs. This has been successfully demonstrated by \citet{2022NatAs...6..350S} with the development of ILT calibration strategies. A 7.4 $\mathrm{deg}^2$ image of the Lockman Hole at 144\,MHz was produced, and its angular resolution is substantially increased from $6\arcsec$ to $0.3\arcsec$.

Currently, over 90\% of the international baseline data for the LoTSS has been recorded. While processing individual sources in the field of view is computationally relatively inexpensive \citep{2022A&A...658A...1M}, producing a multi-degree-scale field of view (FOV) at sub-arcsecond resolution is estimated to take 250,000 core hours \citep{2022NatAs...6..350S}. While developments are being made to reduce this cost, it remains a significant challenge for the capacity of existing computing facilities, particularly when batch image processing is required for survey purposes. This challenge is amplified by the fact that LoTSS alone comprises 3168 pointings, and more pointings would be necessary for a $0.3\arcsec$ resolution survey due to the decreased FOV. Considering this computational bottleneck and the significant disparity between the achievable resolutions, ranging from the highest attainable $0.3\arcsec$ to the standard LoTSS resolution of $6\arcsec$, it is imperative to explore intermediate resolutions.

The ELAIS-N1 field is originally one of a few fields covered by the European Large Area Infrared Space Observatory Survey (ELAIS) \citep{2000MNRAS.316..749O} in the infrared. Since the ELAIS survey, multi-wavelength surveys have been conducted that cover the ELAIS-N1 field, including the Chandra ELAIS deep X-ray survey \citep{2003MNRAS.343..293M} in X-ray, the Medium Deep Survey \citep{2016arXiv161205560C} and the Hyper-SuprimeCam Subaru Strategic Program (HSC-SSP) survey \citep{2018PASJ...70S...8A} in optical, the Galaxy Evolution Explorer (GALEX) survey \citep{2005ApJ...619L...1M, 2007ApJS..173..682M} in ultraviolet, the UKIRT Infrared Deep Sky Survey (UKIDSS) \citep{2007MNRAS.379.1599L}, the Spitzer Extragalactic Representative Volume Survey (SERVS) \citep{2012PASP..124..714M}, and the SIRTF Wide-Area Infrared Extragalactic Survey (SWIRE) \citep{2003PASP..115..897L} in infrared. At radio wavelengths, the ELAIS-N1 field has been covered in multiple large-area radio surveys in a frequency range from 38 MHz to 4.85 GHz. Some of these surveys at lower frequencies include the Cambridge Survey of Radio Sources catalogue at 151 MHz (6C; \citealt{1990MNRAS.246..256H}) and at 38 MHz (8C; \citealt{1995MNRAS.274..447H}), the Very Large Array (VLA) Low-frequency Sky Survey (VLSS) at 74 MHz \citep{2007AJ....134.1245C}, the all-sky TIFR GMRT Sky Survey (TGSS) by the Giant Metrewave Radio Telescope (GMRT) at 150 MHz \citep{2017A&A...598A..78I} and the Westerbork Northern Sky Survey (WENSS) at 325 MHz \citep{1997A&AS..124..259R}. There have also been radio surveys specifically targeting the ELAIS-N1 field, a comparison among those in terms of the area covered, root mean square (RMS) noise and frequency are presented in \citet{2021A&A...648A...2S} (2021, see Figure 1). As one of the four LoTSS Deep Fields \citep{2023MNRAS.523.1729B}, the ELAIS-N1 field was imaged at a resolution of $6\arcsec$, covering $\sim$68 $\mathrm{deg}^2$ \citep{2021A&A...648A...2S}, reaching a root mean square noise level of $\sim$19$\,\mu \mathrm{Jy}\ {\mathrm{ beam}}^{-1}$ in the central region at a central frequency of 144 MHz. This depth is accomplished by amalgamating observations over $\sim$200 hours instead of a standard 8-hour observation. \citet{2021A&A...648A...3K} compiled a multi-wavelength catalogue where multi-wavelength counterparts of radio
sources are identified. 

These multi-wavelength observations establish the ELAIS-N1 field as one of the most extensively observed extragalactic degree-scale fields. However, none of the published images of the ELAIS-N1 field at low radio frequencies achieved a resolution higher than $5\arcsec$. Therefore, an intermediate resolution (e.g. $\simeq 1\arcsec$) image at 144 MHz provides novel information for comprehensive scientific investigations into individual radio sources, also offering unique insights into source morphology compared to the LoTSS $6\arcsec$ resolution images. For example, a Fanaroff-Riley type I/II radio galaxy with its lobes spanning a few arcseconds would only be resolved at higher resolutions \citep{2024A&A...683A..23D}.

When selecting an intermediate resolution to create an image of the ELAIS-N1 field using the LoTSS data with international baselines, we opt for a resolution around $1\arcsec$ in this study for three primary reasons:
\begin{enumerate}
\item Opting for a $1\arcsec$ resolution offers a practical intermediate measure, providing a sixfold improvement over LoTSS' $6\arcsec$ resolution, and will be computationally cheaper than the highest $0.3\arcsec$ resolution by at least a few times. This choice is instrumental in assessing the computational feasibility of imaging the entire Northern Sky at intermediate resolutions between $0.3\arcsec$ and $6\arcsec$, which is a logical step following the LoTSS survey at $6\arcsec$ resolution. 
\item Imaging at $\sim$1$\arcsec$ resolution will advance the detailed study of radio sources, especially star-forming galaxies and active galactic nuclei (AGN) whose emissions are on a scale of $\sim$1$\arcsec$. \citet{2018A&A...618L...8B} find that $\sim$$48$\% of the 3581 $\mu$Jy star-forming galaxies with redshifts ranging from 0 to 7 are resolved at approximately $1\arcsec$ resolution. The remaining resolved sources, resolved higher than $\sim$$0.6\arcsec$, account for 30.8\% of all sources studied. This suggests that the resolved population of $\mu$Jy star-forming galaxies is primarily dominated by sources resolved at approximately $1\arcsec$. The same study also finds that µJy non-radio-excess AGN (NRX-AGN) within the redshift range of 1.50-7.00 and radio-excess AGN (RX-AGN) within the redshift ranges of 0.0-0.3 and 2-7.00 are resolved at approximately $1\arcsec$.
\item Moreover, producing a $\sim$$1 \arcsec$ resolution image for the ELAIS-N1 field matches with the typical resolution of ground-based optical/infrared telescopes, facilitating integration for further multi-wavelength analysis. For instance, existing optical/infrared observations of the ELAIS-N1 field such as MDS \citep{2016arXiv161205560C} have a point spread function (PSF) of $1.2\arcsec$ in the g-band and approximately $1 \arcsec$ in the r, i, z, and y bands. Similarly, SWIRE \citep{2003PASP..115..897L} features a PSF ranging from $1.6\arcsec$ to $2\arcsec$ across wavelengths from 3.6 $\mu$m to 8 $\mu$m, while SERVS \citep{2012PASP..124..714M} has a PSF of $1.7\arcsec$ at 3.6 $\mu$m and  4.5 $\mu$m. These observations have previously been utilised for crossmatching with the LOFAR Deep ELAIS-N1 Field at a resolution of $6\arcsec$. By providing an ELAIS-N1 field image at a resolution of $\sim$$1 \arcsec$, as presented in this work, the potential for more accurate multi-wavelength analysis, including crossmatching, is improved.
\end{enumerate}
Although we have selected a resolution of $\sim$$1 \arcsec$ for this study, it's important to note that other intermediate resolution options are also viable. The imaging approach demonstrated in this paper can be applied to produce images at other intermediate resolutions using the LoTSS survey data with international baselines.

In this paper, we present a $2.5 \times 2.5$ $\mathrm{deg}^2$ image of the ELAIS-N1 field at a resolution of $1.2\arcsec \times2\arcsec$ with a median noise of $\approx$80 $\mu \mathrm{Jy} {\mathrm{ beam}}^{-1}$, and provide its catalogue after careful cross-matching with existing catalogues at radio wavelengths. Advancements have been made in the imaging approach following the creation of the $0.3\arcsec$ resolution image\citep{2022NatAs...6..350S}, particularly in the area of self-calibration. Therefore, these updates are disseminated alongside the image produced. The most up-to-date workflow for making such intermediate-resolution images is outlined. This paper is organised as follows. Section 2 describes the 8-hour ILT observation of the ELAIS-N1 field used in this work. In Section 3, we present the data reduction procedures with a special focus on selecting suitable calibrators within the field to correct for the direction-dependent effects (DDEs). In Section 4, the final image and catalogue are presented after the quality assessment in terms of the flux density scale and positional offsets. Section 5 discusses the source detectability and computational cost of the image we produced. Finally, Section 6 summarises and concludes the work.
                                            
\section{Observations}

On 26 November 2018, the ELAIS-N1 field was observed using LOFAR’s HBAs for a total of 8 hours at frequencies ranging from 120 to 168\,MHz. 3C\,295 was observed prior to the observation for 10 minutes as the primary calibrator. An overview of the observation parameters is given in Table \ref{Table: observation}. This observation used 51 stations including 24 Dutch core stations, 14 remote stations and 13 international stations, resulting in a maximum baseline of 1980.46 km. The layout of LOFAR stations is detailed in \citet{2013A&A...556A...2V} and \citet{2022A&A...658A...1M}. At a frequency of 144 MHz, the core, remote, and international stations have respective fields of view of approximately 12.26 $\mathrm{deg}^2$, 7.90 $\mathrm{deg}^2$, and 3.65 $\mathrm{deg}^2$ \footnote{More information about the LOFAR imaging capabilities and sensitivity can be found at \url{https://science.astron.nl/telescopes/LOFAR/LOFAR-system-overview/observing-modes/LOFAR-imaging-capabilities-and-sensitivity}}. This observation was taken in HBA dual inner mode where only the inner 24 tiles of the 48 in the remote stations are used in order to mimic the core stations, therefore the FoV of the remote stations is the same as the core stations. However, the full 96 tiles in each ILT station are used. An overview of the observation parameters is given in Table \ref{Table: observation}. 

\begin{figure}
   \centering
   \includegraphics[width=0.5\textwidth]{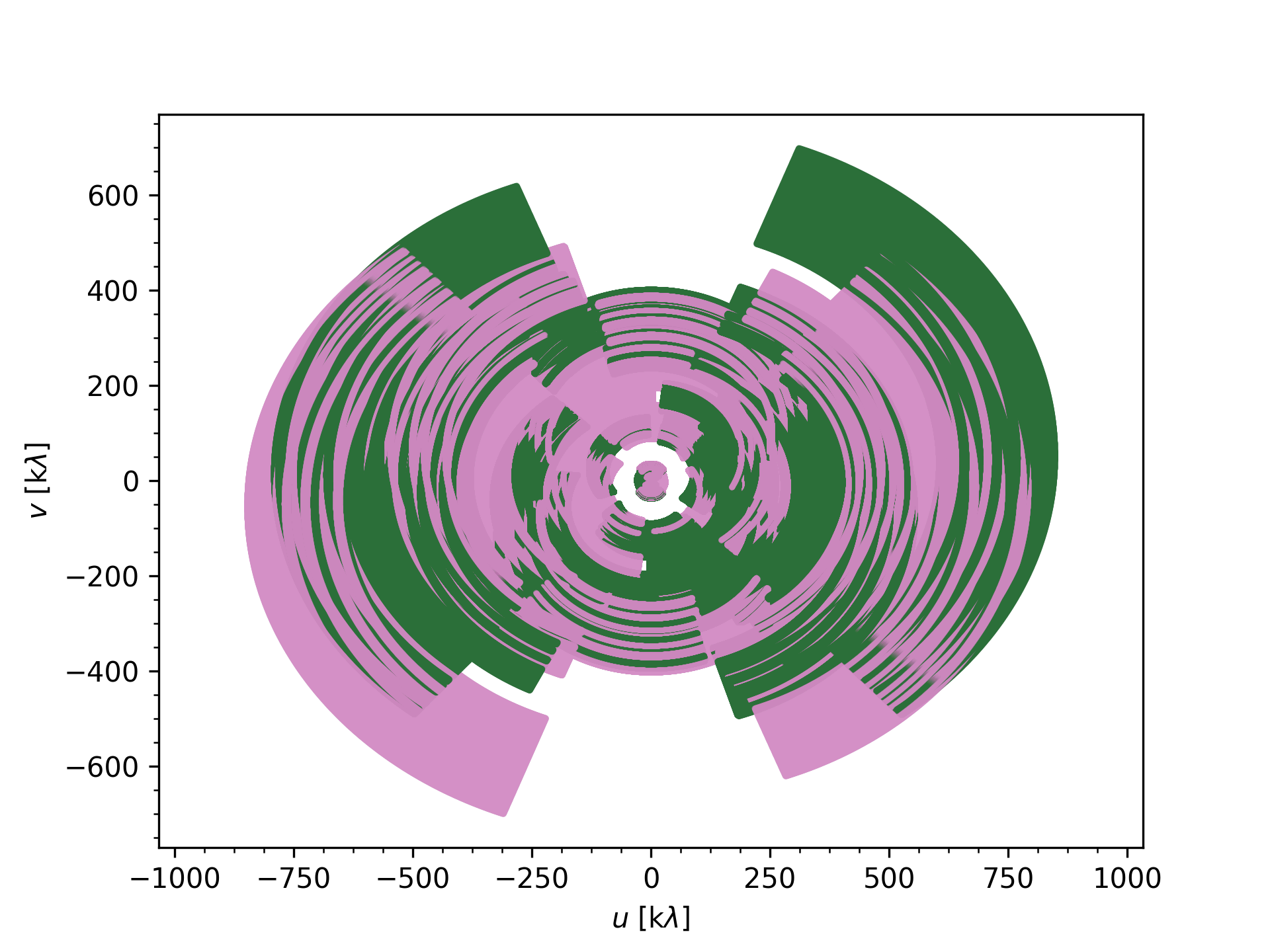}
   \caption{The $uv$-coverage for the ELAIS-N1 field observation at 120-168 MHz. The maximum baseline for this observation is about 2000 km. To improve readability, only one in ten $uv$ points in time and one in 40 $uv$ points in frequency are plotted. The plot depicts symmetric $uv$ points due to conjugate visibilities, where the two colours represent these symmetric $uv$ points.}\label{Fig:uvplot}
   \end{figure}

Figure \ref{Fig:uvplot} plots the $uv$-coverage of this observation, only one in ten $uv$ points in time and one in 40 $uv$ points in frequency are plotted. There are fewer baselines in the range between 40 $k\lambda$ to 90$k\lambda$, due to a scarcity of stations between the Dutch and German HBA locations. 

\begin{table}[]
\centering
\caption{International LOFAR telescope HBA observation parameters}
\label{Table: observation}
\resizebox{\columnwidth}{!}{%
\begin{tabular}{p{0.45\linewidth}p{0.6\linewidth}}
\hline
\noalign{\smallskip}
Observation IDs          & L686956 (3C\,295) \newline                                                                          L686962 (ELAIS-N1)   
\\
Pointing centres         & $14^{h}11^{m}20.4^{s}$+$52^{d}12^{m}10.08^{s}$ (3C\,295)  \newline                              $16^{h}11^{m}00^{s}$+$54^{d}57^{m}00^{s}$ (ELAIS-N1)                                                                 \\
Observation date         & 2018 Nov 26                                                                                                                                                                           \\
Total on-source time     & 10 min (3C\,295)\newline
                           8h (ELAIS-N1)                                                                                                                                                                \\
Correlations             & XX, XY, YX, YY                                                                                                                                                                        \\
Sampling mode            & 8 bit                                                                                                                                                                                 \\
Sampling clock frequency & 200 MHz                                                                                                                                                                               \\
Frequency range          & 120-187 MHz                                                                                                                                                                           \\
Used frequency range          & 120-168 MHz                                                                                                                                                                           \\
Used bandwidth                & 48 MHz (ELAIS-N1,3C\,295)                                                                                                                                                       \\

Bandwidth per SB         & 195.3125 kHz                                                                                                                                                                          \\
Channels per SB          & 64                                                                                                                                                                                    \\

Stations                 & 51 total\newline 13 International\newline 14 remote\newline 24 core (48 split\protect\footnotemark)                                                                             \\ 
\noalign{\smallskip}
\hline
\end{tabular}
}
\end{table}
\addtocounter{footnote}{0}
\footnotetext{Each core station is split into two substations.}
%
   
   \mycomment{
   $uv$-coverage for the used ELAIS-N1 field observation at 120-168 MHz. The maximum baseline is 1514 km (or 853 $k\lambda$). In this figure, only one in ten $uv$ points in time and one in 40 $uv$ points in frequency are plotted. The two colours show the symmetric $uv$ points due to conjugate visibilities.
   In plot_uv.py:
   >>> print "The longest baseline is {n} km".format(n = max(np.abs(udata))/1000)
   The longest baseline is 1505.88712571 km
   >>> print "The largest v value is {n} km".format(n = max(np.abs(vdata))/1000)
   The largest v value is 1242.13115963 km
   >>> print "The longest baseline is {n} km".format(n = max(np.sqrt(vdata**2 + udata**2)/1000))
   The longest baseline is 1514.77838778 km
   >>> wave = 2.998e8/ max(freq_col)
   >>> print "The longest baseline is {n} km".format(n = max(np.sqrt(vdata**2 + udata**2)/1000))
   The longest baseline is 853.517879597 k lambda
   }

\begin{figure*}
\centering
\includegraphics[width=0.45\textwidth]{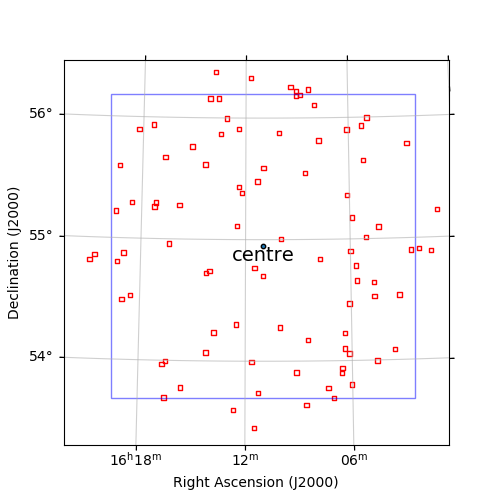}
\includegraphics[width=0.45\textwidth]{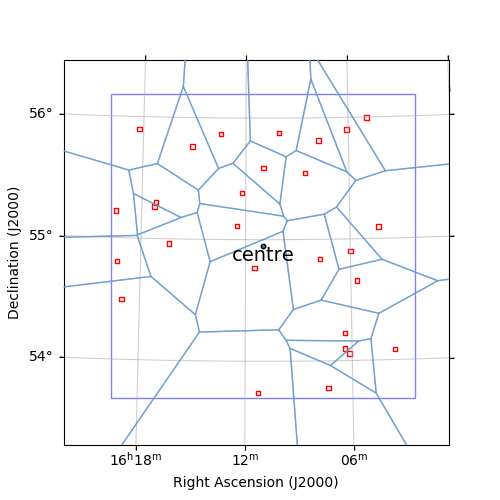}
\caption{The boundary of the ELAIS-N1 field image made is defined by the outer blue square. The positions of direction-dependent calibrators are denoted by red small squares, while the pointing centre is marked by a blue solid dot. The left panel shows all self-calibration calibrator candidates, and the right panel displays the selected 28 calibrators. Each blue polygon in the right panel contains one calibrator, which is applied to correct the direction-dependent effects (DDEs) of the corresponding region.}
\label{fig:selfcal_dirs}
\end{figure*}

\section{Data Processing: calibration and imaging}

The procedures for making the final calibrated and deconvolved $2.5  \times 2.5$ $\mathrm{deg}^2$ image from our 8-hour ELAIS-N1 field observation can be divided into 4 major steps:

\begin{enumerate}
    \item Calibration of all Dutch stations
    \item Direction-independent calibration for international stations 
    \item Direction-dependent calibration for international stations 
    \item Making an image at a resolution of $\sim$$1\arcsec$ from the calibrated data
\end{enumerate}
The $2.1$ TB ILT data we began with has a time and frequency resolution of 2 seconds and $12.207\ \mathrm{kHz}$, respectively, referred to as the `original data'.

\subsection{Calibration of Dutch stations}\label{sec:3.1}

We started the standard LOFAR HBA data reduction and calibration process using \texttt{Prefactor} \citep{2016ApJS..223....2V, 2016MNRAS.460.2385W, 2019A&A...622A...5D}\footnote{\url{https://github.com/LOFAR-astron/prefactor}}. 
This pipeline consists of two parts: a \textit{calibrator} pipeline and a \textit{target} pipeline. Target observations are usually bookended by short observations of a flux-density calibrator for redundancy. The solutions from the flux-density calibrator preceding this target observation were sufficient, so we did not use the following calibrator observation. The calibrator pipeline utilised these observations to rectify three significant systematic effects: the phase offsets between the X and Y polarisations, station bandpasses, and clock offsets between stations. These corrections were derived for all stations, including international ones. This was done using a $\sim$10-minute observation of 3C\,295 and applying a source model appropriate for sub-arcsecond imaging, set to the \citet{2012MNRAS.423L..30S} flux scale. Subsequently, the target pipeline applies these corrections, averages the data to an integration time of $8$ seconds and a frequency channel width of $48.828\ \mathrm{kHz}$, and removes the international stations. The Faraday rotation was then corrected first, using RMextract \citep{2018ascl.soft06024M}. Finally, a phase calibration was performed using the TGSS sky model \citep{2017A&A...598A..78I}. This gives a direction-independent bulk correction of the ionospheric corruptions for the Dutch stations.

Direction-dependent calibration to correct remaining DDEs across the FOV was subsequently performed with the ddf-pipeline\footnote{\url{https://www.github.com/mhardcastle/ddf-pipeline}} \citep{2019A&A...622A...1S, 2021A&A...648A...1T}. This provided a high-quality $6\arcsec$ resolution model of the field, spanning approximately $8.3 \times 8.3$ $\mathrm{deg}^2$ in size, which will be used for source subtraction in the subsequent step. 

\subsection{Direction-independent Calibration of the international stations}

The LOFAR-VLBI pipeline \citep{2022A&A...658A...1M} was employed to conduct the direction-independent calibration of the international stations. This started with applying the solutions obtained by \texttt{Prefactor} to the original data. Next, bright and compact sources within the ELAIS-N1 field were selected, using the \textit{Long-Baseline Calibrator Survey} (LBCS, \citet{2016A&A...595A..86J, 2022A&A...658A...2J}, as candidate in-field calibrators, also referred to as delay calibrators, for direction-independent calibration of the international stations. From these LBCS candidates, ILTJ160607.63+552135.5 was taken as the in-field calibrator due to its compact nature and high SNR of the calibration solutions. The data was then phase-shifted to the in-field calibrator position, and all core station visibility data were phased up into a single superstation, which suppresses flux from neighbouring sources on all baselines involving the superstation. After that, the data was averaged in both time and frequency to further suppress the other sources in the FOV beyond the in-field calibrator.

Self-calibration was used to calibrate the chosen in-field calibrator. As a starting model, a point-source model was used. After an initial round of phase calibration against the starting model, an updated model was created after each iteration. Three iterations of phase calibration were followed by 5 iterations of phase and amplitude calibration on multiple short time scales (8 seconds for the phases and an order of 30 minutes for amplitude) following the calibration strategy outlined by \citet{2022NatAs...6..350S}, for a total of eight iterations.

After that, the $6\arcsec$ resolution model produced and described in \ref{sec:3.1} was subtracted outside of the international station's central FOV ($2.5  \times 2.5$ $\mathrm{deg}^2$). The aim of this step is to suppress the interference from sources outside the target region when performing the wide-field imaging, which is intensified on Dutch stations due to their field of view being twice as large, leading to less primary beam suppression of distant sources compared to international stations. 

As a final step, solutions derived from the
self-calibration procedure described above on the in-field calibrator is applied to the data.

\subsection{Direction-dependent calibration for international stations}\label{subsec: selfcal}

\begin{figure}
\centering
\includegraphics[width=0.5\textwidth]{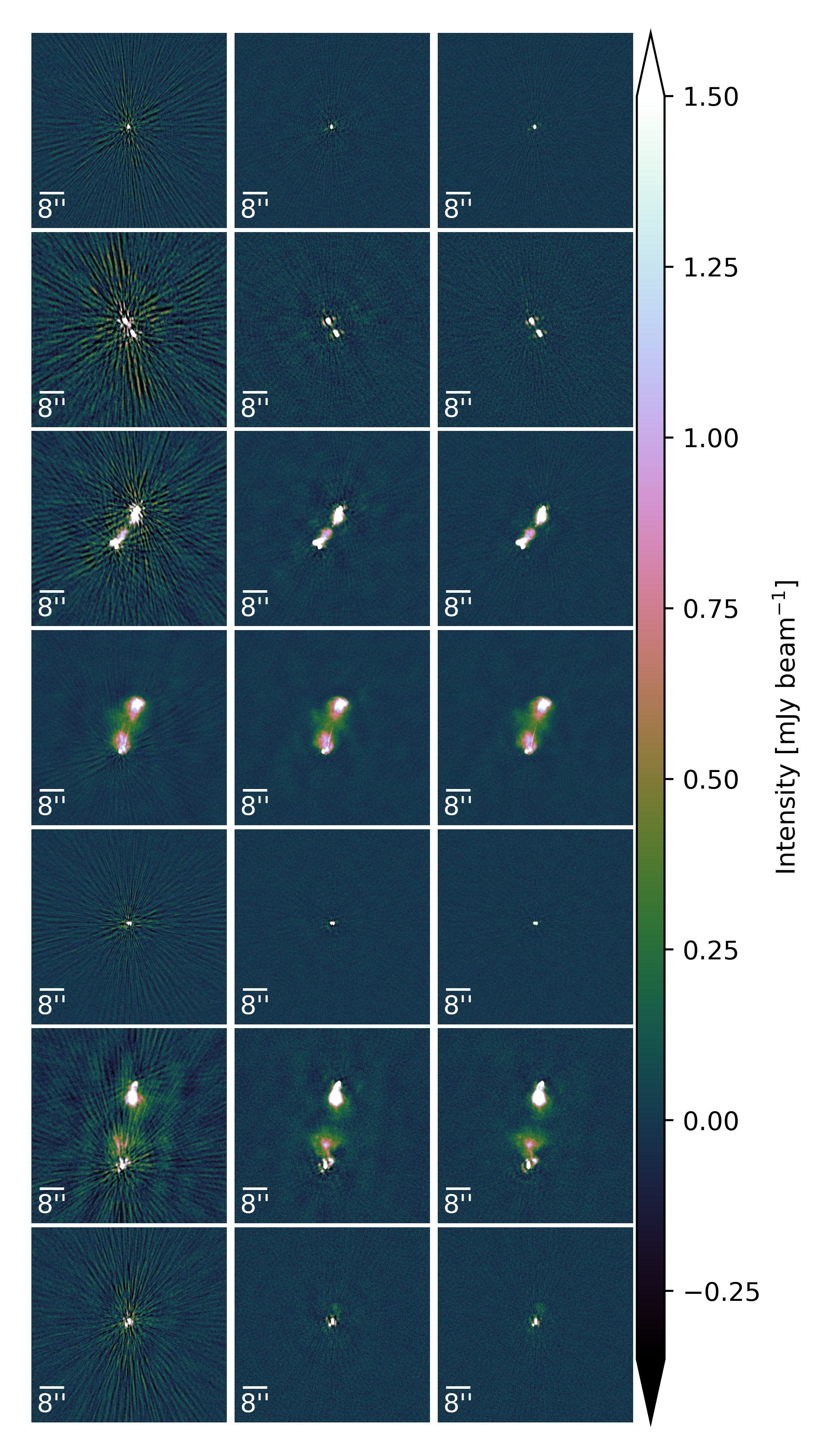}
\caption{Each row of images showcases the iterative self-calibration process of one calibrator in correcting direction-dependent effects. The three columns correspond to the calibrator's images with 0, 2, and 4 iterations, representing no correction, dTEC-only correction, and dTEC and amplitude/phase correction, respectively. Each image has a size of $64 \times 64$ arcsec and contains 1600 pixels by 1600 pixels.}
\label{fig:selfcal_example}
\end{figure}

\begin{figure}
\centering
\includegraphics[width=0.5\textwidth]{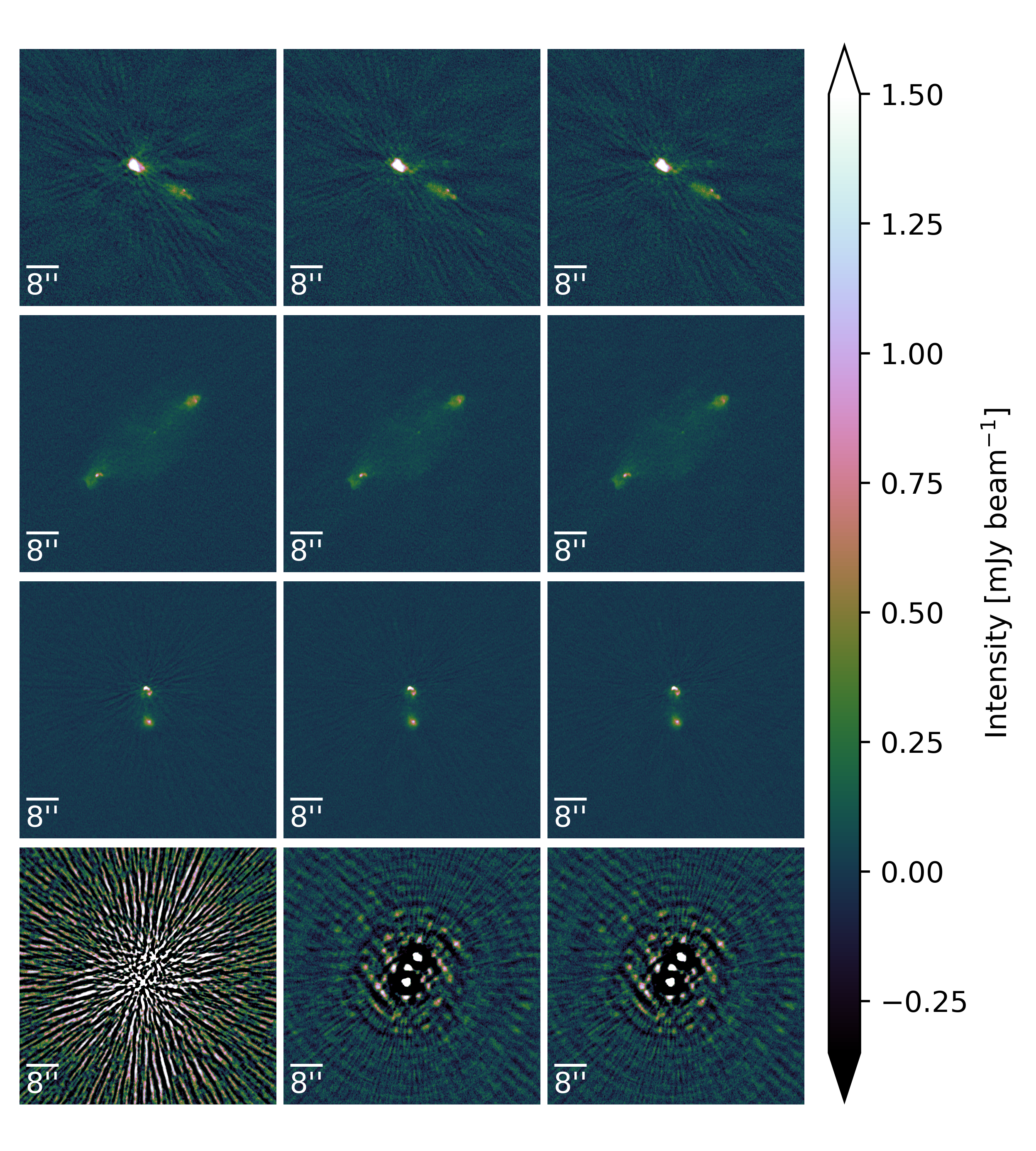}
\caption{The self-calibration of the four calibrator candidates in which correcting direction-dependent effects (DDEs) was unsuccessful. The three columns show the four candidates' images with 0, 4, and 8 iterations of self-calibration, representing no correction, dTEC-only correction, and dTEC and amplitude/phase correction, respectively. Each image has a size of $64 \times 64$ arcsec and contains 1600 pixels by 1600 pixels.}
\label{fig:selfcal_bad_example}
\end{figure}

\begin{figure*}
   \centering
   \includegraphics[width=\textwidth]{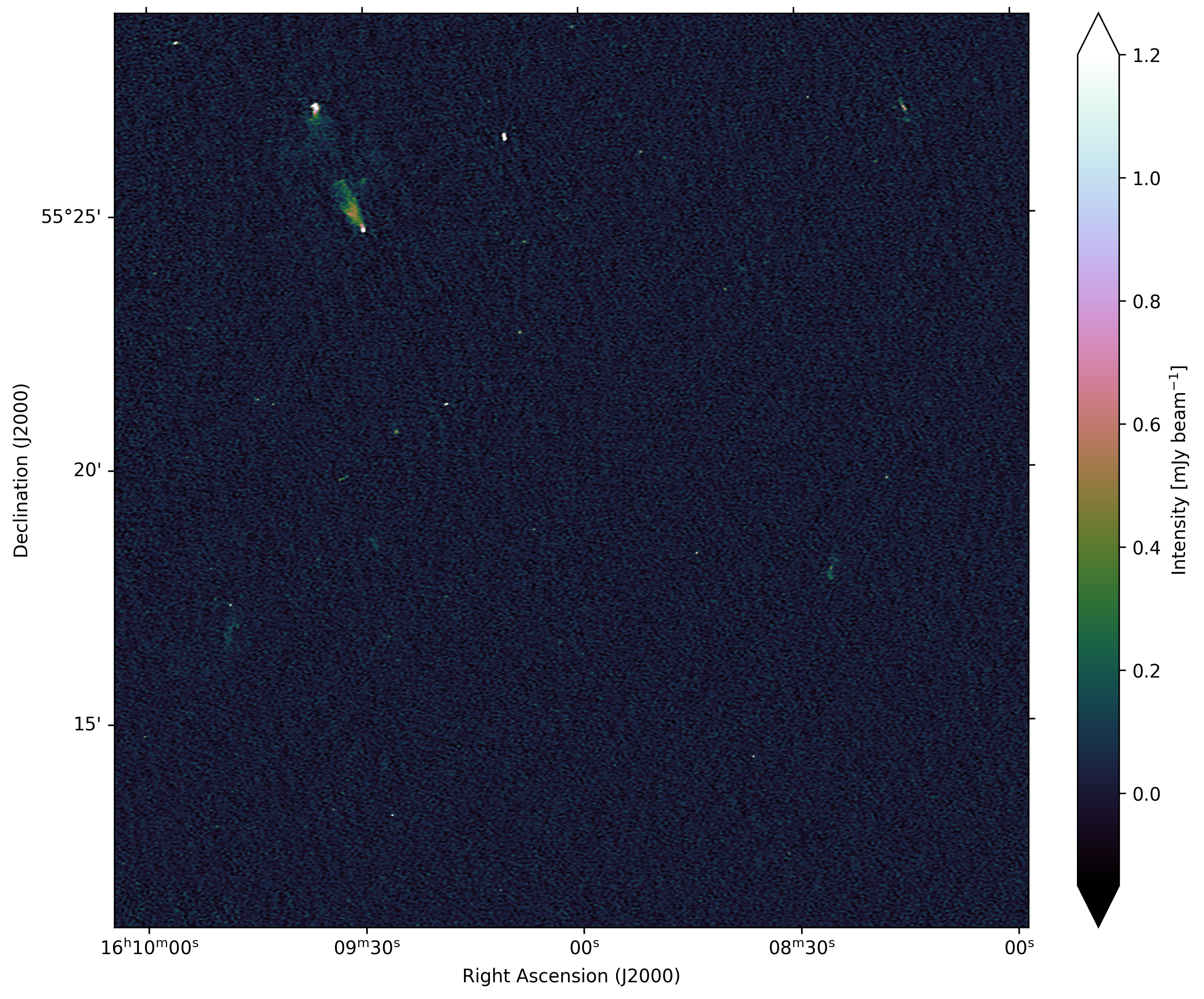}
   \caption{This zoom-in image depicts a central region in the ELAIS-N1 field, which was created using an 8-hour international LOFAR telescope observation at 120-168 MHz. This region is centred at ($16^h09^m01.36^s$, $55^m19^m56.7^s$) and is $0.3 ~\mathrm{deg} \times 0.3~\mathrm{ deg}$ in size, with a resolution of $1.2\arcsec$ $\times$ $2\arcsec$. The colour bar represents the flux density from -$2\sigma$ to $20\sigma$, where $\sigma = 0.122$ mJy beam$^{-1}$ is the approximate RMS noise in this region of the image.}\label{Fig:zoomin_image1}%
\end{figure*}

In order to map the ELAIS-N1 field, it is imperative to mitigate the residual direction-dependent effects (DDEs), which are predominantly induced by the ionosphere. In accordance with the DDE calibration methodology outlined by \citet{2022NatAs...6..350S}, we initially identified 93 potential DDE calibrators from the ELAIS-N1 LOFAR Deep field radio catalogue \citep{2021A&A...648A...2S}. 
The selection criterion for these candidates was based on their peak intensities, which were required to be greater than $25$ mJy per beam. Subsequently, 93 distinct data sets were generated by phase-shifting at the position of each respective DDE calibrator and averaged. Each data set was averaged down to a 1-minute time interval and a $\sim$0.4 MHz (195.3125 kHz $\times$ 2) frequency interval, with an aim of introducing smearing within a few arcminutes from the phase centre to suppress the inference of other sources. It also reduces subsequent processing time, as no solution intervals below this resolution are required after the in-field calibration. The left panel of Figure \ref{fig:selfcal_dirs} illustrates the RA-DEC distribution of these DDE calibrator candidates. 

For each of the 93 data sets, we undertook an iterative self-calibration procedure, as described by \citet{2021A&A...651A.115V} to correct for their total electron content (TEC) values of international stations. This was achieved through the \texttt{facetselfcal.py}\footnote{\url{https://github.com/rvweeren/LOFAR_facet_selfcal}} Python script \citep{2021A&A...651A.115V}, which employs the Default Preprocessing Pipeline \citep[\texttt{DPPP,}][]{2018ascl.soft04003V} and \texttt{WSCLEAN} \citep{2014MNRAS.444..606O, 2017MNRAS.471..301O}. To be more specific, each self-calibration iteration generated one solution. Each solution was used to correct the differential TEC values of the dataset using a short solution interval ($\sim$1\,min), and optionally correct the phase and amplitude (i.e., complex gains) on longer timescales ($>20$\,min) for candidates with a higher flux density to correct the inaccuracy in the LOFAR beam model. 
We applied the solution obtained at each iteration to produce an image of size $64 \times 64$ arcsec, with a pixel size of $0.04\arcsec$. The images, centred on the respective calibrator candidates, were generated using \texttt{WSCLEAN} with a resolution of $0.3\arcsec$, which is the full resolution for ILT. Figure \ref{fig:selfcal_example} demonstrates the iterative DDE calibration process for seven calibrator candidates after completing the 0th, 2nd, and 4th iteration respectively, representing no correction, dTEC-only correction, and dTEC and complex gain correction, respectively. The images in each column of the figure depict the gradual reduction of calibration artefacts resulting from DDEs after the completion of the 2nd and 4th self-calibration iterations.

Not all calibrator candidates yield self-calibration solutions that effectively correct for nearby DDEs. There are candidates that do not have enough flux density on the longer baselines to be successfully self-calibrated. Some candidates exhibit unaltered noise levels even after undergoing several iterations, while others have inadequate dynamic ranges (defined as the absolute value of the ratio between the maximum and minimum intensity values of a given image). Four examples are shown in Figure \ref{fig:selfcal_bad_example}, where the first, second, and fourth (from top to bottom) calibrator candidates result in a dynamic range of less than 20 after 8 rounds of calibration iterations, and the second candidate provided the lowest dynamic range of the three ($\sim$10). The third candidate illustrated in the figure achieved a good dynamic range of around 36 at Iteration 2; however, after that, the dynamic range dropped and converged at Iteration 4, with a noise level reduction of less than 8\% compared to its 0th Iteration, when no calibration had been applied. 

Our experience with including a few such calibrator candidates for this direction-dependent calibration step has shown that their ineffectively corrected self-calibration solutions can introduce significant calibration errors in the final image product. Therefore, we developed specific selection criteria to identify calibrators that can effectively address DDEs, as follows:
\begin{enumerate}
    \item During each iteration, the image of the selected calibrator must exhibit an increasing trend in the image dynamic range and a decreasing trend in noise level until convergence.
    \item The minimum image dynamic range of the calibrator must exceed a threshold value. In our case, the threshold was set to $28$.
    \item The application of self-calibration should result in a reduction percentage in noise level that exceeds a given threshold. In our study, the threshold was set to $8\%$ as compared to when no self-calibration was applied.
\end{enumerate}
It is worth noting that the chosen threshold values of $28$ and $8\%$ were based on our experimental findings and may vary for different observations and calibrator sets. Further investigation is currently in progress to explore the theoretical basis for the selection. Furthermore, the minimum number of selected calibrators and their distribution within a given region size requires further attention.

As a result, we selected 28 calibrators. To further validate our selection, we visually inspected the images of each chosen calibrator to confirm that the DDE effects surrounding them had been effectively removed. There were two pairs of calibrators, each separated by $3.14$ arcminutes and $2.28$ arcminutes respectively, that were selected. Since we phase-shifted the visibilities to centre at each respective DDE calibrator, the phase-up operation starts suppressing the influence of other sources within a radius of approximately $\sim 1\arcmin$ (see \cite{2022A&A...658A...1M}; Figure 5).

For each of the 28 data sets, we selected one solution from multiple iteration solutions based on a combination of dynamic range and noise level. The noise level here is used for selection as amplitudes are rescaled to 1 during amplitude self-calibration steps, so the flux density scale of the calibrators is well-constrained throughout multiple iterations. To be more specific, we select the solution that achieves both the lowest noise level and the highest dynamic range on the image of the calibrator to which this solution has been applied. For example, if the 6th iteration produces the lowest noise level and highest dynamic range, we would select the 6th solution. In cases where the highest dynamic range precedes the iteration with the lowest noise level, we would prioritise selecting the iteration with the lowest noise level. Therefore, if the image produced by the 4th iteration demonstrated the highest dynamic range, but the lowest noise level occurred in the 8th iteration, we would choose the solution from the 8th iteration. 

In Figure \ref{fig:selfcal_dirs}, the right panel showcases the distribution of the selected DDE calibrators, marked by red boxes. The self-calibration solution chosen for each of these calibrators will be applied to their respective surrounding areas (or facets), marked by blue boundaries. No other self-calibration solutions were used during the final imaging process.

\section{Results}

\subsection{Image result}
\begin{figure*}
   \centering
   \includegraphics[width=\textwidth]{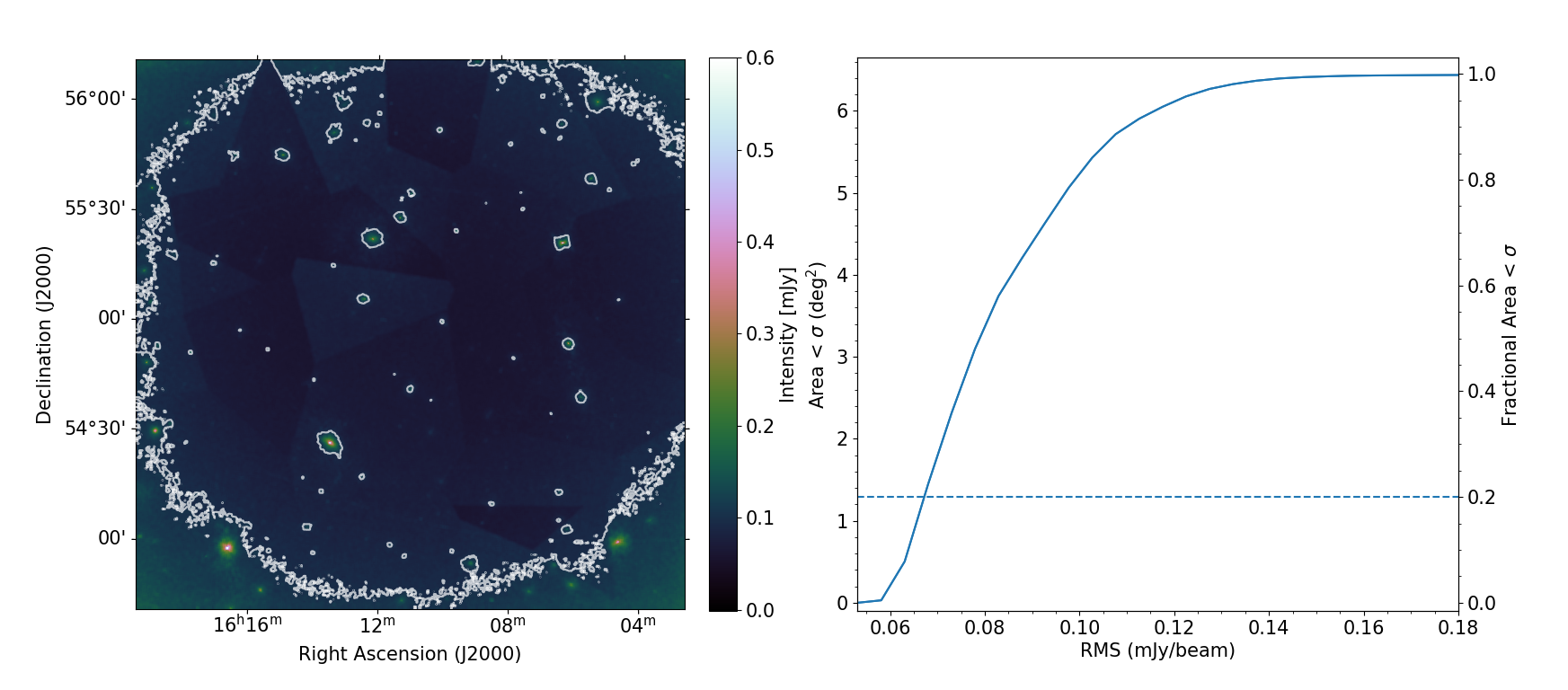}
   \caption{The left panel of this figure shows the RMS noise image of the ELAIS-N1 field with a contour level at 0.1 mJy $\mathrm{beam}^{-1}$ overlaid. The contours at the positions of bright sources indicate that their direction-dependent effects (DDEs) are not completely eliminated. The right panel depicts the cumulative ratio of pixels in the RMS image against RMS noise values. The $x$-axis represents the RMS noise value, while the $y$-axis shows the percentage of image pixels with RMS values greater than or equal to a given RMS noise value, along with its corresponding physical area.}\label{Fig:rms_contour}%
\end{figure*}

\begin{figure}
   \centering
  \includegraphics[width=0.5\textwidth]{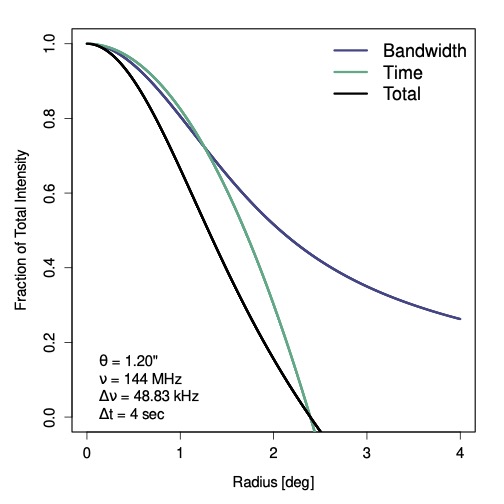}
   \caption{Effects of bandwidth and time smearing: changes in the fraction of initial total intensity for a point source as a function of its distance from the pointing centre. The calculation is based on Eqns. (18-43) and (18-24) of \citet{1999ASPC..180..371B}}\label{Fig:smearing}%
\end{figure}

Before imaging, the data were averaged to an integration time of 4 seconds and a frequency channel width of 48.828 kHz. To apply the DDE correction solutions from the 28 calibrators, we used the \texttt{facet-imaging} mode of \texttt{WSCLEAN} and generated a $2.5 \times 2.5$ $\mathrm{deg}^2$ image of the ELAIS-N1 field, covering an area of 6.45 $\mathrm{deg}^2$. We limited the $uv$ data to be larger than $80\lambda$. Our imaging process employed Briggs' weighting (robust = $-1.5$), auto-masking, the multi-scale CLEAN deconvolution algorithm \citep{4703304, 2017MNRAS.471..301O}, and \texttt{WSCLEAN}'s wide-field imaging module \texttt{wgridder} \citep{2021A&A...646A..58A, 2022MNRAS.510.4110Y}. The resulting image has a size of 22\,500 by 22\,500 pixels, with a pixel size of $0.4\arcsec$ and a taper size of $1.2\arcsec$. 

The final primary-beam-corrected image at $1.2\arcsec \times2\arcsec$  resolution can be accessed online \footnote{\url{https://home.strw.leidenuniv.nl/~wwilliams/LoTSS_1arcsec}}. Its positional offsets and flux density scale have been corrected, and the procedures for these corrections will be discussed in Section \ref{sec: catalogue}. To illustrate the resolution, quality, and a range of sources in the field, Figure \ref{Fig:zoomin_image1} displays a $0.3 \times 0.3$ $\mathrm{deg}^2$ area of the final image, centred at ($16^h09^m01.36^s$, $55^m19^m56.7^s$), which is 28.5 arcminutes from the image phase centre ($16^h11^m00^s$, $54^m57^m00^s$). 

We also generated an RMS noise map from the flux- and astrometric-corrected image using the source finder package \texttt{PyBDSF} \citep{2015ascl.soft02007M} described in Section \ref{sec: catalogue}. The resulting RMS noise image is shown on the left side of Figure \ref{Fig:rms_contour}, with a contour level corresponding to a value of 0.1 mJy ${\mathrm{ beam}}^{-1}$ overlaid to the false colour image. The RMS noise map indicates a noticeable increase in noise level from the image's
centre to its edge due to the primary beam correction. The contours surrounding some of the brightest sources suggest that their DDEs have not been completely removed. In the right plot of Figure \ref{Fig:rms_contour}, we show the cumulative (fractional) area mapped as a function of RMS noise. To make this plot, we extracted the noise value from each pixel of the RMS image and binned the number of pixels for given RMS noise value ranges. As seen in this plot, the inner 20\% of the image has a noise level below 0.068 mJy ${\mathrm{ beam}}^{-1}$. 

Figure \ref{Fig:smearing} provides information on the amount of bandwidth and time smearing when imaging at a resolution of $1.2\arcsec$, with time and frequency resolution of 4 seconds and 48.83 kHz, respectively. We considered a compromise between the amount of smearing and imaging speed when selecting the averaging settings. For general reference, 20\% losses occur at a radius of 0.74 degrees, while 50\% losses occur at a radius of 1.30 degrees. 

\begin{figure}
   \centering
  \includegraphics[width=0.5\textwidth]{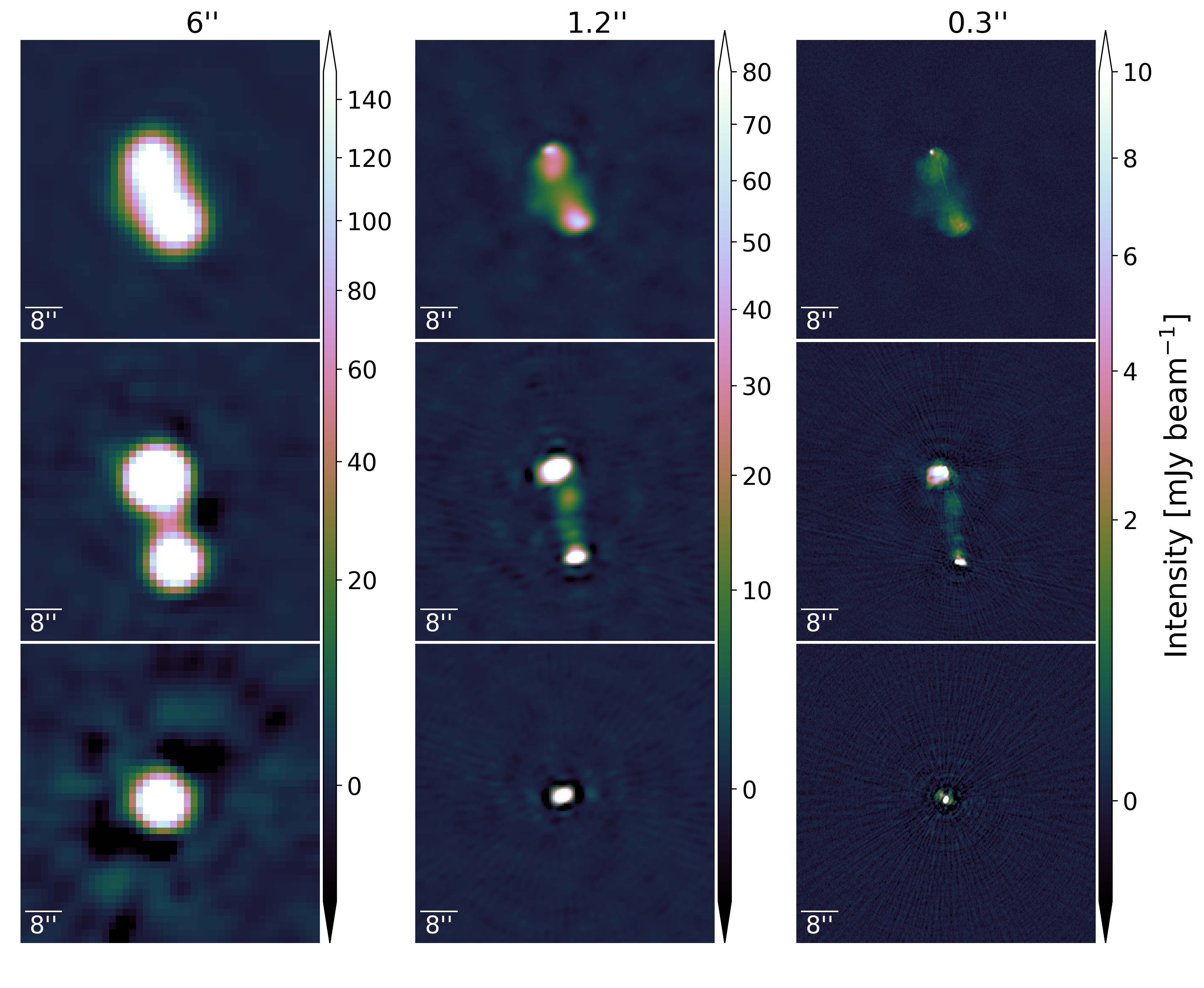}›
   \caption{Three sources imaged at three different resolutions, namely $6\arcsec$, $1.2\arcsec$, and $0.3\arcsec$. Left: cutouts extracted from the LOFAR deep field image at a resolution of $6\arcsec$; centre: cutouts from our $1.2\arcsec$ resolution image, which is presented in this paper; right: $0.3\arcsec$ resolution images generated using \texttt{WSCLEAN} during the self-calibration process.}\label{Fig:comparison_image}%
\end{figure}

In Figure \ref{Fig:comparison_image}, we show three sources imaged at resolutions of $6\arcsec$, $1.2\arcsec$, and 0.3$\arcsec$ respectively. These sources were selected from our catalogue of 28 DDE calibrators. The $6\arcsec$ resolution images were cutouts from the LOFAR ELAIS-N1 deep field image \citep{2021A&A...648A...2S}, whereas the $0.3\arcsec$ resolution images were obtained during the self-calibration procedure detailed in Section \ref{subsec: selfcal}. Figure \ref{Fig:comparison_image} reveals that these sources, which appeared compact in the $6\arcsec$ resolution image, exhibit significant levels of resolved emission at higher resolutions. Consequently, the high-resolution images provide more intricate and informative details about the sources. We also selected 40 extended sources whose peak flux is larger than 2 mJy ${\mathrm{beam}}^{-1}$ and a FWHM of their major axis larger than 7.2 arcseconds to display in the Appendix \ref{appendix}.

\subsection{Radio catalogue}\label{sec: catalogue}
\begin{figure*}
   \centering
   \includegraphics[width=0.45\textwidth]{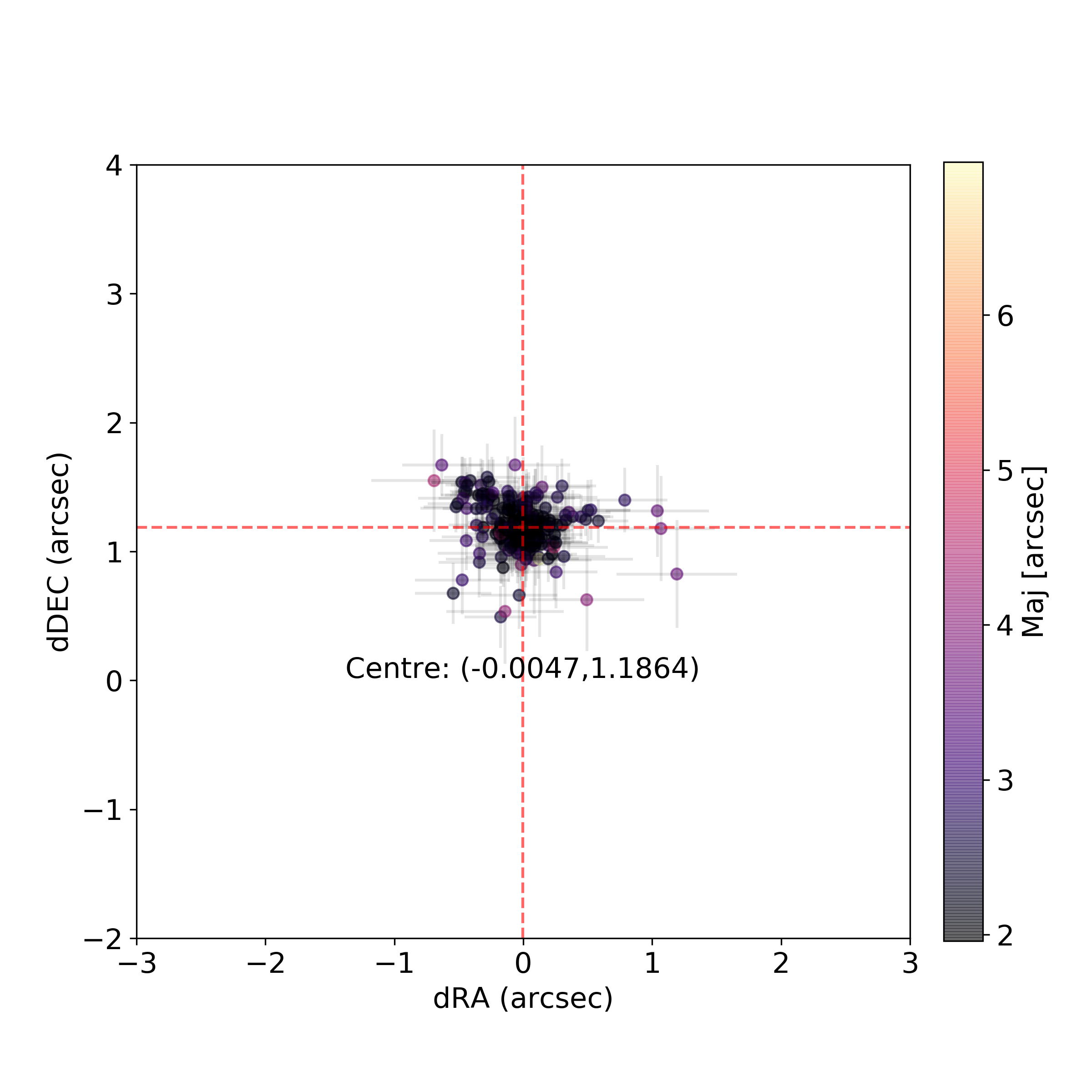}
   \includegraphics[width=0.45\textwidth]{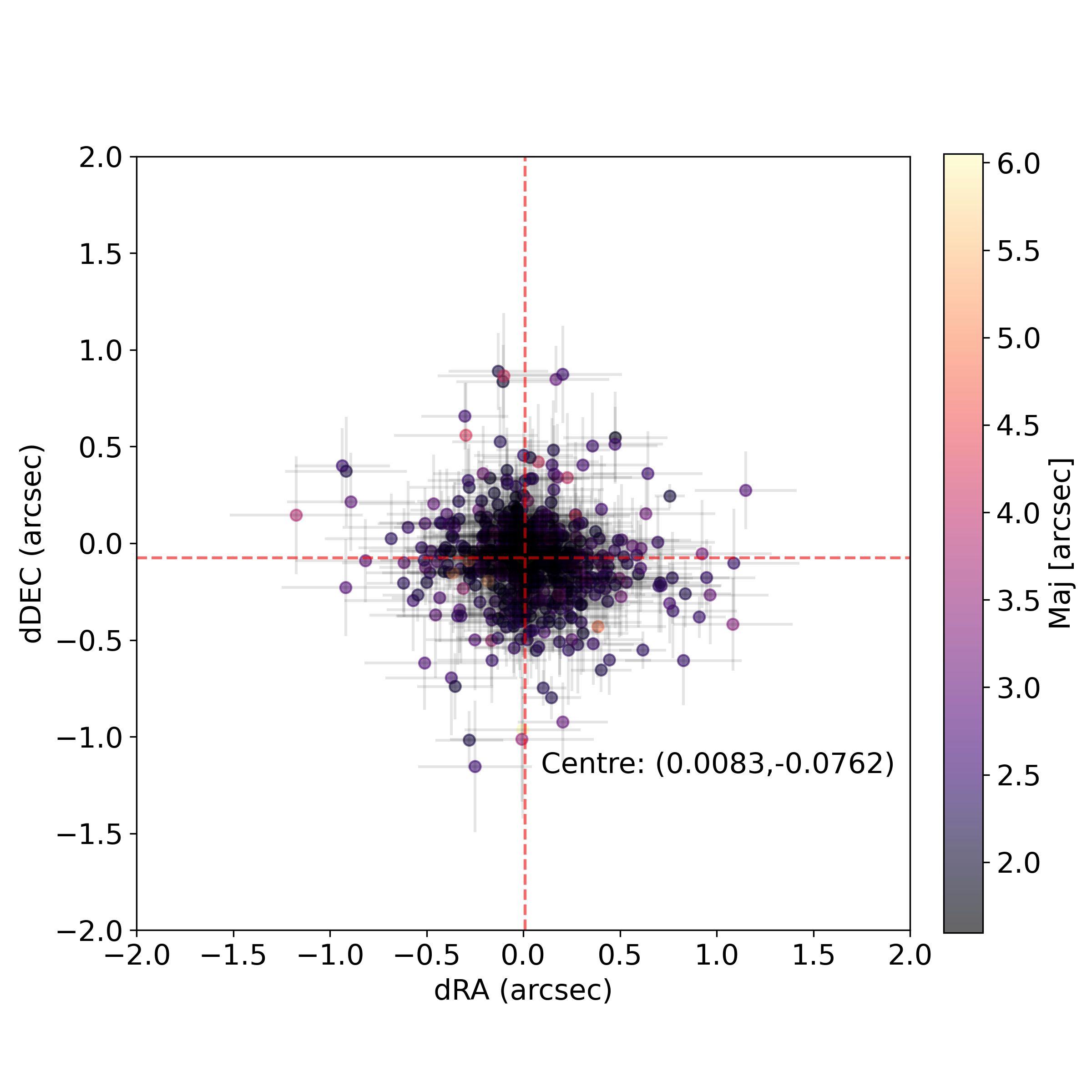}
   \caption{Left: positional offsets of selected cross-matched sources between two catalogues, one is extracted from the $6\arcsec$ resolution LOFAR deep image of the ELAIS-N1 field, and the other one is obtained from our current image at a resolution of $1.2\arcsec$. The median values of the offset are $\mathrm{dRA} = -0.0047\arcsec$ ($\sigma= 0.0623\arcsec$) and $\mathrm{dDEC}= 1.1864\arcsec$ ($\sigma$ = 0.0336$\arcsec$). Right: Positional offsets of selected cross-matched sources from two catalogues, one is the optical catalogue of the ELAIS-N1 field, and the other one is our corrected radio catalogue. The median values of the offset are $\mathrm{dRA}$ = 0.0083$\arcsec$ ($\sigma$ = $0.0817\arcsec$) and $\mathrm{dDEC} = -0.0762\arcsec$ ($\sigma$ = 0.0605$\arcsec$). The near-to-zero median values validate the quality of the position offset correction. The error bar of each source is taken from the source extraction output of package \texttt{PyBDSF}. }\label{Fig:dra_ddec_6deep}
\end{figure*}

To generate a preliminary radio catalogue from our ELAIS-N1 image, we employed the source extraction package \texttt{PyBDSF}, which fits sources using one or more Gaussians. The parameters used in \texttt{PyBDSF} were taken from the HBA deep fields settings (see Appendix C of \citet{2021A&A...648A...2S}). \texttt{PyBDSF} detected 3\,797 sources. In addition to generating the radio catalogue, \texttt{PyBDSF} also generated an RMS noise map as displayed in the left panel of Figure \ref{Fig:rms_contour}, as well as fitted Gaussian and residual maps. 

\begin{figure*}
   \centering
  \includegraphics[width=0.45\textwidth]{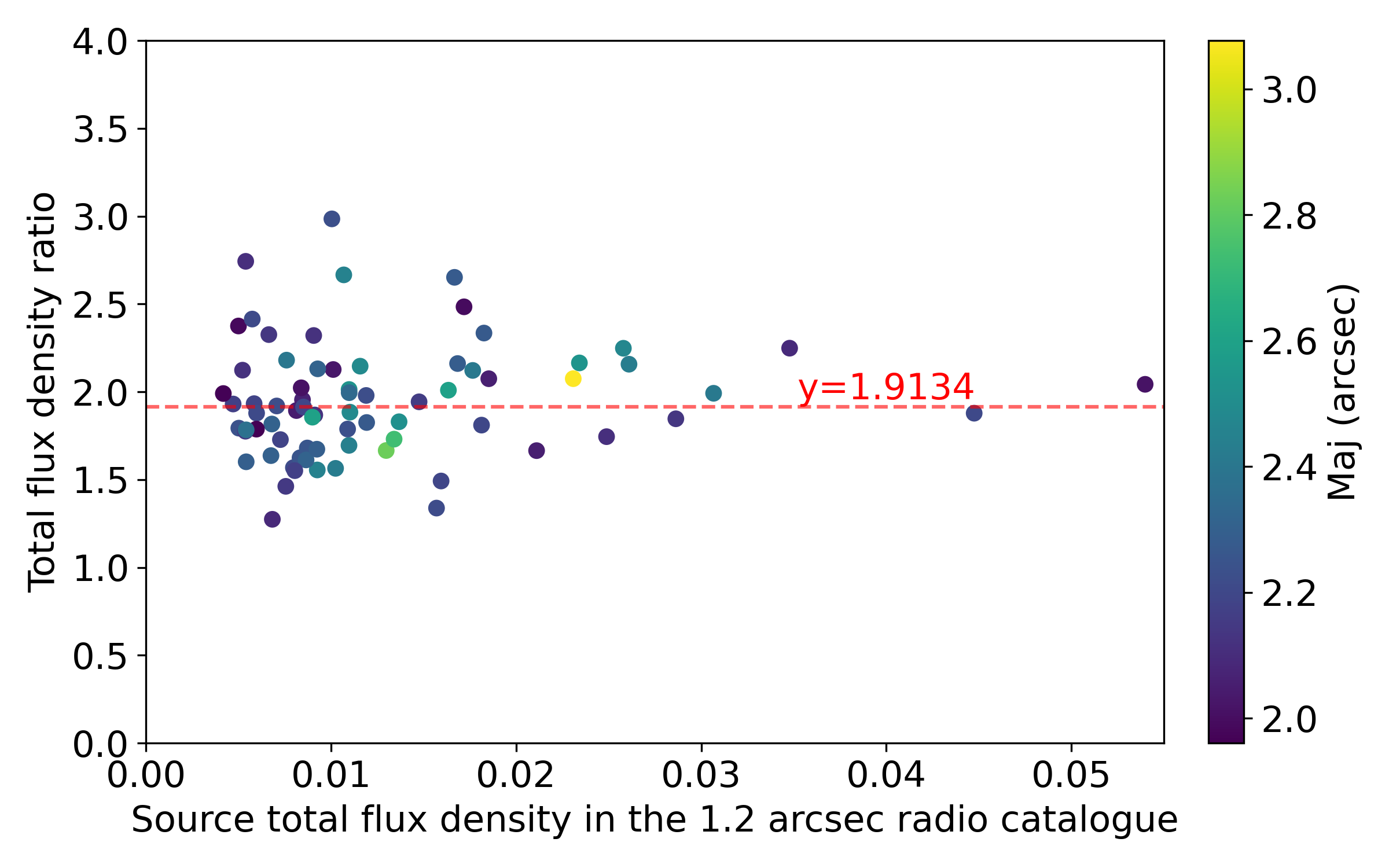}
  \includegraphics[width=0.45\textwidth]{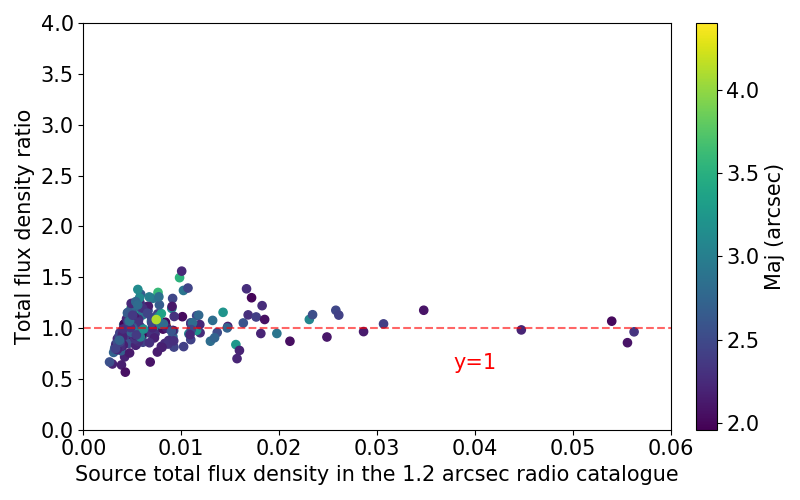}
   \caption{Left: The flux density ratio of 77 compact sources between our $1.2\arcsec$ resolution radio catalogue and the $6\arcsec$ resolution deep field catalogue is presented. The median value of 1.9134 is used as the flux scaling factor. Right: The flux scaling factor is applied to correct the flux densities of 223 compact sources in our $1.2\arcsec$ resolution catalogue. As a result, the flux density ratios between the corrected flux densities and their corresponding values in the $6\arcsec$ resolution deep field catalogue are observed to be scattered around 1.}\label{Fig:flux_ratio}
\end{figure*}

\begin{figure}
   \centering
  \includegraphics[width=0.45\textwidth]{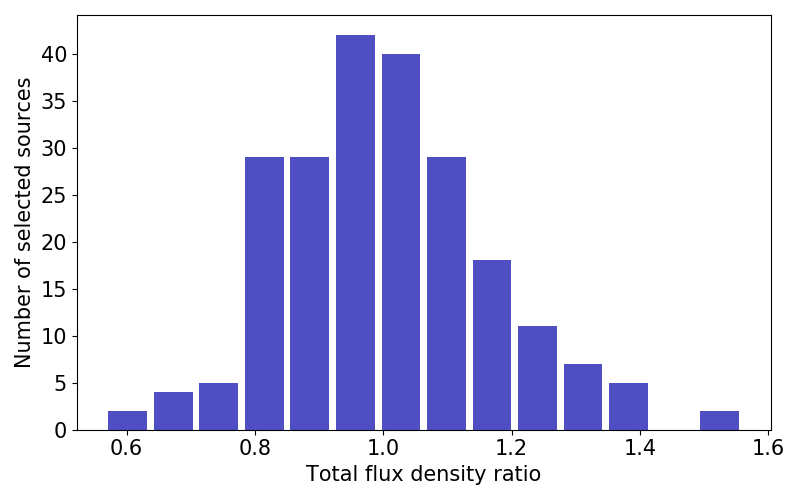}
   \caption{The histogram presents the flux density ratios of 223 selected compact sources, with the distribution observed to be centred around 1. This confirms the validity of the obtained scaling factor.}\label{Fig:flux_hist}
\end{figure}

\subsubsection{Astrometric precision}\label{subsec: astrometry}

As the positions of our sources were extracted from our $1.2\arcsec$ resolution ELAIS-N1 image, any phase calibration errors in making this image could result in source position offsets. To address this issue, we cross-matched the radio catalogue produced by \texttt{PyBDSF} with the LOFAR $6\arcsec$ resolution ELAIS-N1 deep field radio catalogue \citep{2021A&A...648A...2S} using \texttt{TOPCAT} \citep{2005ASPC..347...29T}, allowing for a maximum positional error of $6\arcsec$. The $6\arcsec$ resolution radio catalogue was extracted from the LOFAR $6\arcsec$ resolution deep HBA image of the ELAIS-N1 field \citep{2021A&A...648A...2S}, which has undergone examination through multi-wavelength source associations and cross-identifications with multiple optical observations \citep{2021A&A...648A...3K}, so it serves as a high-quality benchmark for our catalogue. 
As a result, 2\,990 sources were cross-matched. 

Compact and bright sources tend to have more accurate positions because it is easier for the source extraction package to measure their positions with lower uncertainties, as opposed to extended or less bright sources. Subsequently, we selected 231 compact and bright sources from the cross-matched catalogue to assess the astrometric accuracy of our image. The selection criteria were as follows: 1) each selected source has a recorded flux density larger than 2 mJy in both catalogues; 2) only one Gaussian component is fitted; and 3) The FWHM of the major axis is smaller than $7.2\arcsec$ (1.2 times the resolution of the deep image). The selection of 7.2’’ instead of $6\arcsec$ allows sources to be slightly larger than the beam, as suggested in \citet{2022A&A...659A...1S}. These criteria are applied to both the LOFAR $6\arcsec$ ELAIS-N1 deep field radio catalogue and the new catalogue generated from the $1.2\arcsec$ resolution image. 

In Figure \ref{Fig:dra_ddec_6deep}, we illustrate the positional offsets in both right ascension (RA) and declination (DEC) for these 231 selected sources. The RA offset of a selected source is defined as the difference between its RA values in the $6\arcsec$ resolution deep field image and our $1.2\arcsec$ resolution image, while the declination offset is denoted as dDEC in a similar manner. The median of the position offset is $\mathrm{dRA}= -0.0047\arcsec$  ($\sigma =0.0623 \arcsec$) and $\mathrm{dDEC}= 1.1864 \arcsec$  ($\sigma= 0.0336 \arcsec$). The right ascension offset is approximately zero, while the declination offset is more substantial. This pronounced declination offset is largely attributed to the in-field calibration procedure, where the in-field calibrator is selected from the LBCS survey. Its position is taken from the WENSS survey, which has a resolution of $54\arcsec \times 54\arcsec$ and positional accuracy of 1.5$\arcsec$ for strong sources \citep{1997A&AS..124..259R}. Consequently, an astrometric correction is necessary for our $1.2\arcsec$ resolution image.

Additionally, the colour bar of the scattering points in Figure \ref{Fig:dra_ddec_6deep} represents the FWHM of the major axis of each source, demonstrating that larger sources tend to have a larger positional offset. 

To verify the accuracy of the position offset correction, we performed an additional round of cross-matching. We first applied the median values of the positional offset to both RA and DEC axes of our $1.2\arcsec$ resolution image to correct its astrometric precision and used \texttt{PyBDSF} to extract sources from the updated $1.2\arcsec$ resolution image. Secondly, we cross-matched the resulting catalogue with the optical source catalogue based on a combination of various optical/IR observations with a resolution range from $0.9\arcsec$ to $1.72\arcsec$ \citep{2021A&A...648A...3K}. 

Following the initial crossmatching, we applied the following selection criteria to this subset of cross-matched sources in the radio catalogue only: 1) the SNR of the source, calculated by its peak flux density divided by its average background RMS value taken from the `Isl\_rms' column of the catalogue generated by \texttt{PyBDSF}, is larger than 10; 2) ensuring the radio source's flux density is greater than 2 mJy; and 3) each radio source is only fitted with a single Gaussian component with a major axis FWHM smaller than $7.2\arcsec$. As a result, a final selection of 560 sources was obtained. 

The median values of the positional offset for these selected sources were $\mathrm{dRA}= 0.0083 \arcsec (\sigma =0.0817 \arcsec)$ and $\mathrm{dDEC}= -0.0762$ $(\sigma= 0.0605 \arcsec)$. While we could have applied more constraints to the selection to reduce the number of samples, the fact that the median values of these 560 samples were close to zero, and both the median and variance were of the same order, this validates that the position offset has been accurately corrected.

We then investigated the facet-dependent variation in the positional offsets, as detailed in Appendix \ref{appendix:2}. We found that the variations across the field are small and well below the resolution of our image. Therefore, we have decided to apply only this one astrometric correction.

\subsubsection{Flux density scale}\label{subsec:fluxscale}

To ensure the accurate flux densities of our catalogued sources, we started with the flux density of our infield calibrator, ILTJ160607.63+552135.5. This calibrator has a flux density of 0.2352 Jy in the 6-arcsecond resolution LOFAR deep field radio catalogue. However, in our extracted radio catalogue, its flux density was found to be 0.4115 Jy. This discrepancy indicates the necessity for a flux scaling correction, with the flux density scaling factor estimated to be around 2. Since the unaveraged visibility data was phase-shifted to the position of this calibrator before the infield calibration, the smearing effects at the infield calibrator have been minimised to a negligible level.

\begin{table*}[htbp]\footnotesize
  \centering
  \caption{The catalogue's example entries}
  \tiny
  \begin{tabularx}{\textwidth}{lcXcXccccc}
    \hline
    Source & RA & $\sigma_{\mathrm{RA}}$ & DEC & $\sigma_{\mathrm{DEC}}$ & $S_i$ & $S_p$ & a & b & $\phi$      \\
     Name& (deg) & (arcsec) & (deg) & (arcsec) & (mJy)& (mJy/bm) & (arcsec) & (arcsec) &  (deg)   \\
    (1) & (2) & (3)  & (4)  & (5)  & (6)  & (7)  & (8) & (9)  & (10)        \\
     \hline
ILTJ161212.32+552303.7&243.05132&0.03&55.38437&0.01&$2572.2\pm4.57$&$489.2\pm0.33$&$19.4\pm0.06$&$3.4\pm0.01$&$12.4\pm0.23$\\   ILTJ161900.64+542937.1&244.75265&0.07&54.49364&0.02&$1786.5\pm8.11$&$114.0\pm0.36$&$23.2\pm0.16$&$5.0\pm0.03$&$9.6\pm0.56$\\   ILTJ160538.36+543922.7&241.40982&0.01&54.6563&0.01&$1291.6\pm2.32$&$197.9\pm0.18$&$13.6\pm0.04$&$3.0\pm0.01$&$32.0\pm0.19$\\        ILTJ160600.00+545405.7&241.5&0.04&54.90157&0.02&$1171.2\pm6.15$&$52.7\pm0.29$&$14.4\pm0.1$&$5.4\pm0.04$&$20.6\pm0.63$\\  ILTJ161640.39+535812.9&244.16831&0.18&53.97025&0.08&$966.2\pm13.64$&$119.6\pm0.89$&$16.5\pm0.45$&$5.1\pm0.13$&$21.3\pm2.29$\\   ILTJ160454.75+555949.7&241.22812&0.01&55.99715&0.0&$883.9\pm1.9$&$312.2\pm0.29$&$6.3\pm0.02$&$1.9\pm0.0$&$146.0\pm0.17$\\       ILTJ161507.57+554540.6&243.78155&0.0&55.76128&0.0&$718.5\pm0.93$&$469.4\pm0.25$&$2.2\pm0.0$&$1.6\pm0.0$&$110.8\pm0.08$\\     ILTJ161331.29+542718.1&243.38038&0.02&54.45503&0.03&$400.7\pm3.59$&$96.0\pm0.58$&$5.0\pm0.07$&$2.4\pm0.03$&$62.8\pm1.12$\\    ILTJ160435.47+535936.8&241.14778&0.01&53.99355&0.01&$326.4\pm4.0$&$81.8\pm0.41$&$4.5\pm0.03$&$2.4\pm0.01$&$102.8\pm0.52$\\    ILTJ161002.79+555242.7&242.51163&0.0&55.87853&0.0&$305.6\pm0.85$&$238.1\pm0.23$&$2.1\pm0.0$&$1.4\pm0.0$&$114.5\pm0.13$\\
. &&&&&&&&&\\
. &&&&&&&&&\\
. &&&&&&&&&\\
ILTJ160940.75+544733.4&242.41977&0.1&54.79262&0.12&$0.5\pm0.12$&$0.4\pm0.07$&$1.9\pm0.33$&$1.4\pm0.17$&$141.9\pm20.95$\\
ILTJ160922.94+551101.3&242.34558&0.16&55.1837&0.09&$0.5\pm0.12$&$0.4\pm0.06$&$2.1\pm0.4$&$1.4\pm0.17$&$106.2\pm17.27$\\
ILTJ160734.92+550224.0&241.89551&0.11&55.04&0.1&$0.5\pm0.12$&$0.4\pm0.06$&$1.8\pm0.3$&$1.4\pm0.18$&$124.7\pm24.83$\\
ILTJ160855.59+551408.8&242.23163&0.14&55.23577&0.11&$0.5\pm0.13$&$0.4\pm0.06$&$1.9\pm0.34$&$1.6\pm0.24$&$101.6\pm41.12$\\
ILTJ161504.20+545155.7&243.76751&0.13&54.86546&0.07&$0.5\pm0.11$&$0.5\pm0.06$&$1.9\pm0.31$&$1.3\pm0.14$&$109.6\pm16.52$\\
ILTJ161537.90+550150.4&243.90793&0.11&55.03066&0.14&$0.5\pm0.13$&$0.4\pm0.06$&$1.8\pm0.32$&$1.6\pm0.27$&$180.0\pm73.24$\\
ILTJ161449.99+552339.8&243.7083&0.11&55.39438&0.11&$0.5\pm0.12$&$0.4\pm0.06$&$1.8\pm0.29$&$1.5\pm0.23$&$129.5\pm50.48$\\
ILTJ160806.66+550507.2&242.02776&0.17&55.08533&0.06&$0.5\pm0.11$&$0.4\pm0.06$&$2.2\pm0.4$&$1.2\pm0.13$&$101.6\pm11.53$\\
ILTJ160929.93+550601.7&242.37473&0.18&55.10047&0.13&$0.5\pm0.13$&$0.3\pm0.06$&$2.1\pm0.43$&$1.6\pm0.28$&$106.3\pm37.45$\\
ILTJ160832.28+550450.2&242.1345&0.1&55.08061&0.06&$0.5\pm0.09$&$0.5\pm0.05$&$1.8\pm0.26$&$1.2\pm0.11$&$111.9\pm11.74$\\
  \hline
  \end{tabularx}
  \label{tab:catalogue}
  \\
  \raggedright
  \footnotesize\textbf{Notes:}\\
  (1) Source name\\
  (2,3) Position right ascension (RA), and uncertainty \\
  (4, 5) Position declination (Dec), and uncertainty \\
  (6) integrated flux density and uncertainty \\
  (7) peak flux intensity and uncertainty \\
  (8-10) fitted shape parameters and their corresponding uncertainties: deconvolved major- and minor-axes, and position angle, for extended sources, as determined by \texttt{PyBDSF}.
\end{table*}

To obtain the exact flux density scaling factor, we selected 77 compact and bright sources from the cross-matched catalogue between our $1.2\arcsec$ resolution position-corrected radio catalogue and the $6\arcsec$ resolution LOFAR deep field radio catalogue. The selection criteria were applied to both catalogues as follows: 1) the SNR of the source, calculated by its peak flux density divided by its average background RMS value taken from the `Isl\_rms' column of the catalogue generated by \texttt{PyBDSF}, is larger than 30; 2) ensuring the total flux density is greater than 2 mJy; and 3) each radio source is only fitted with a single Gaussian component with a major axis FWHM smaller than $7.2\arcsec$. As a result, these sources have fewer uncertainties in their flux density measurements than extended or less bright sources.

In the left scatter plot of Figure \ref{Fig:flux_ratio}, we display the total flux density ratio of the chosen compact sources. This ratio is defined as the result of dividing the total flux density of the same source in our $1.2\arcsec$ resolution catalogue by its counterpart in the $6\arcsec$ resolution deep catalogue. Unlike peak intensities, the total intensities are not affected by bandwidth and time smearing. The $x-$axis shows the flux density of each source in our preliminary $1.2\arcsec$ resolution radio catalogue, while the $y-$axis displays their flux density ratios. From these ratios, we calculated a median value of 1.9134, which we subsequently adopted as the flux density scaling factor. This factor aligns coherently with the initial scaling factor estimation we projected during the investigation of the infield calibrator. However, this scaling factor is larger than that derived by \citet{2022NatAs...6..350S} for the Lockman Hole field at sub-arcsecond resolution, which was recorded $1.21 \pm 0.19$. It is worth emphasising that during the Direction-dependent calibration step for international stations outlined in Section \ref{subsec: selfcal}, the amplitudes were normalised to prevent flux-scale drifting. This relatively large scaling factor likely originates from the derived bandpasses for the international stations, which, for the observations listed in this paper, used the complex source 3C\,295. At the time, there was no very accurate model available for the longest baselines, as the source is very resolved.

To validate the obtained scaling factor, we selected a larger group of sources with less restrictive criteria: we relaxed the SNR criterion from above 30 to above 10. As a result, 223 sources were selected. We applied the scaling factor by dividing the flux density of each source in the $1.2\arcsec$ resolution catalogue by 1.9134 and recreated the flux density ratio plot. The resulting plot is displayed in the right panel of Figure \ref{Fig:flux_ratio}. Combined with the histogram shown in Figure \ref{Fig:flux_hist}, we observed that the total flux density ratio of these 223 sources is centred at 1, validating the obtained scaling factor. As no systematic facet-dependent effect was found (see analysis in Appendix \ref{appendix:2}), we applied this flux density scaling factor to our astrometry-corrected $1.2\arcsec$ resolution image through a division. This corrected flux scale follows the flux scale outlined in Section 3.5 of \citet{2021A&A...648A...2S}. 


\subsubsection{Corrected radio catalogue}

After correcting for the position offset and flux scale, we obtained the final ELAIS-N1 image at a resolution of $1.2\arcsec$ with good astrometry and flux measurements. The resulting image is $\sim$1.9 GB in size.

We used the same parameters as before to run \texttt{PyBDSF} on the updated image and detected a total of 3\,921 sources. To ensure the reliability of the detections, we applied three criteria to the detected sources:
\begin{enumerate}
    \item No part of the source should be located outside the image boundaries.
    \item The flux intensity should be above a minimum threshold of $7.5\sigma$, and the peak intensity should exceed $5.5\sigma$, where $\sigma$ is taken from this source's `Isl\_rms' column of the catalogue generated by \texttt{PyBDSF}.
    \item The sources should have counterparts in the deeper $6\arcsec$ LOFAR image to avoid false detections resulting from statistical fluctuations. This means that we are not considering the possibility of any transient source.
\end{enumerate}
As a result, our \texttt{PyBDSF} generated radio catalogue has 2\,263 sources. This catalogue is similar to the Elais-N1 deep field catalogue presented by \cite{2021A&A...648A...2S}.

A sample of this catalogue generated from our $1.2\arcsec$ resolution LOFAR HBA image of the ELAIS-N1 field ($2.5  \times 2.5$ $\mathrm{deg}^2$), showing the brightest and faintest entries, is given in Table \ref{tab:catalogue}. It should be noted that peak intensities are affected by both bandwidth and time smearing, while the total fluxes remain unaffected.

\section{Discussions}

\begin{figure}
   \centering
   \includegraphics[width=0.45\textwidth]{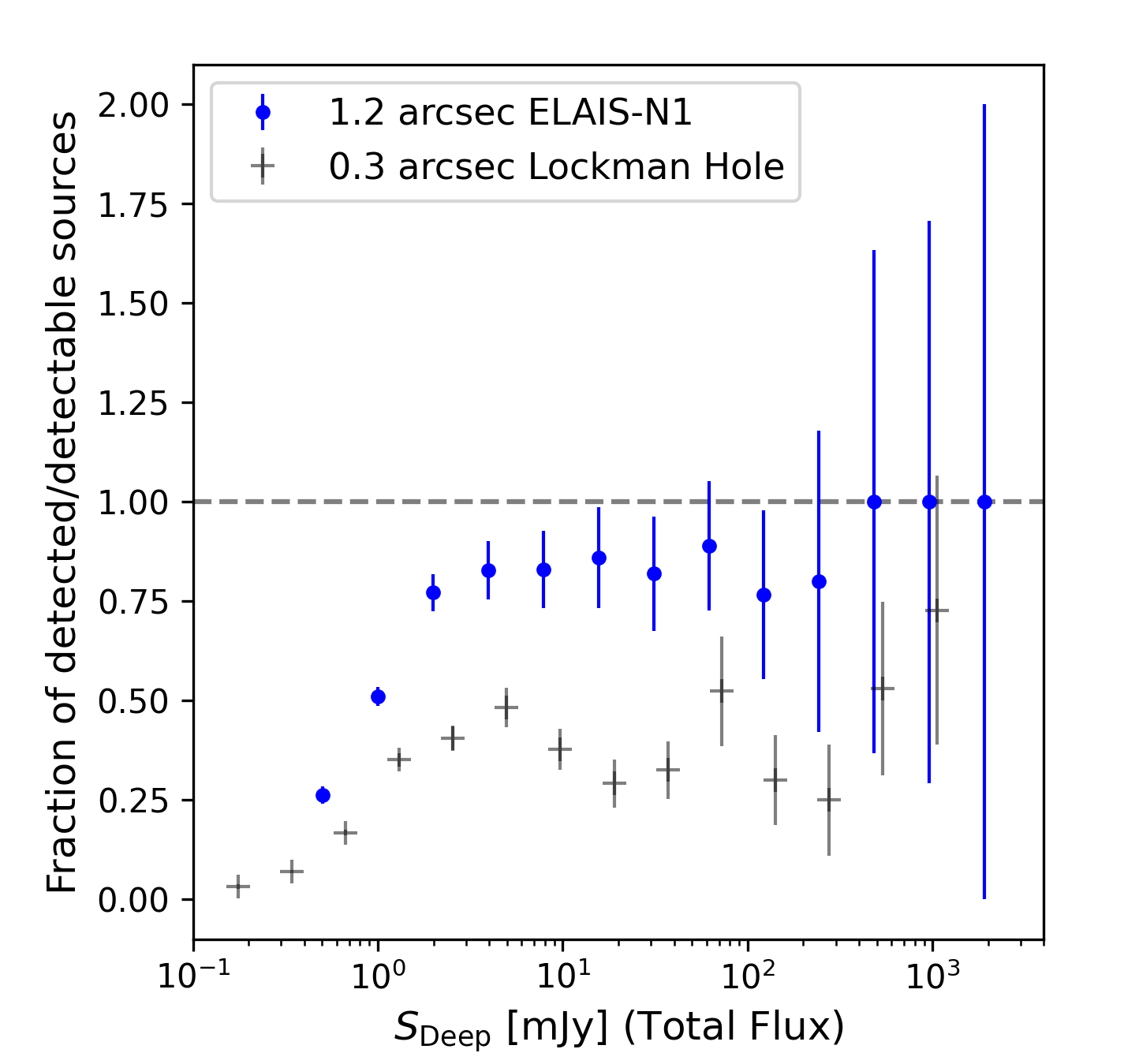}
   \caption{This plot shows the ratio of detected to detectable sources in the $1.2\arcsec$ resolution image, as a function of flux density (from the $6\arcsec$ resolution LoTSS Deep Field catalogue). Sources from the $6\arcsec$ resolution LOFAR deep field ELAIS-N1 catalogue are considered detectable if their peak intensity exceeds 5.5 times the RMS at the same coordinates in our $1.2\arcsec$ resolution ELAIS-N1 image. The blue dots demonstrate this ratio. Uncertainties are estimated using the $\sqrt{N}$ method and propagated accordingly.}\label{Fig:ddetection_ratio}
\end{figure}

\subsection{Discussion on detectability comparison}

We investigated the likelihood that a source in the $6\arcsec$ resolution LOFAR deep field will be detected at higher resolution. This will depend on the amount of flux density which is in compact components and will change as a function of integrated flux density as the population shifts from star-forming galaxies and radio-quiet AGN to radio-loud AGN. 

In Figure \ref{Fig:ddetection_ratio}, the blue solid circles depict the ratios of sources detected in the $1.2\arcsec$ resolution image to those that should be detectable based on their peak intensity in the $6\arcsec$ resolution LOFAR deep field ELAIS-N1 radio catalogue. A source is considered detectable if its peak intensity is greater than 5.5 times the RMS at the same coordinates in our $1.2\arcsec$ resolution ELAIS-N1 image. To analyse the results further, sources are separated into different flux bins based on their total flux density. For each bin, the percentage of detected sources is then calculated. The uncertainties on the plot are estimated using the $\sqrt{N}$ method, which considers the number of sources in each bin, and propagates accordingly. It is expected that bright and compact sources are more easily detectable than dim and extended sources. 

In contrast, the black crosses represent the same ratio of detected to detectable sources, but this data corresponds to the $0.3\arcsec$ resolution wide-field image of the Lockman Hole, which can be referred to Fig. 8 of the study by \citet{2022MNRAS.515.5758M}. Although these are different fields, we can make a general comparison of detection rates between resolutions of $0.3\arcsec$ and $1.2\arcsec$; we note that the fields had the same observational setup and data processing strategy up to the point of imaging. 

It is immediately obvious that more sources are detected in the $1.2\arcsec$ resolution image at all total flux densities, by a factor of $\sim$2, except for the very highest flux bin ($\sim$1 Jy). 
This underscores the importance of the $1.2\arcsec$ resolution image for population studies. We do not observe the 5 mJy dip present in the $0.3\arcsec$ resolution Lockman Hole catalogue, likely due to the increased sensitivity to low-surface-brightness radio emission at arcsecond scales in the $1.2\arcsec$ resolution image. This increased sensitivity allows the $1.2\arcsec$ resolution image to detect more radio emission from the faint population compared to the $0.3\arcsec$ resolution Lockman Hole image, which is more sensitive to compact emission.


\subsection{Discussion on computational cost}

It took $\sim$52\,000 core hours to generate this $1.2\arcsec \times 2\arcsec$ resolution image from the 8-hour LOFAR LoTSS observation data for the ELAIS-N1 field, which is nearly five times quicker than imaging at sub-arcsecond resolution (which requires approximately 250\,000 core hours; \citet{2022NatAs...6..350S}). 

The approximate core hour distribution for this process includes $\sim$2\,000 for calibrating all Dutch stations using \texttt{Prefactor}, $\sim$10\,000 for direction-dependent calibration for Dutch stations with the ddf-pipeline \citep{2019A&A...622A...1S, 2021A&A...648A...1T}, $\sim$7\,000 performing direction-independent calibration for international stations, $\sim$10\,000 for subtracting the $6\arcsec$ resolution model, and another $\sim$10\,000 for completing direction-dependent calibration for international stations.

Subsequent to these calibrations, the imaging step consumes $\sim$13\,000 core hours of computational resources, and it is also one of the most memory-intensive steps. It takes around 6 days to produce the final $1.2\arcsec$ resolution image from the fully-calibrated 8-hour LOFAR observation, which can run on a single compute node. The node used for imaging consisted of 512 GB RAM and dual 24-core Dual Intel Xeon Gold 5220R with hyper-threading. This is at least one order of magnitude cheaper in terms of core hours compared to making a sub-arcsecond resolution image \citep{2022NatAs...6..350S}, and also faster in terms of wall time by a factor of a few. Consequently, even a modestly-sized computing infrastructure could handle large-scale imaging at a manageable cost and within a reasonable time frame, enable studies of sources in patches of sky at $\sim 1\arcsec$ resolutions. Although a resolution of $\sim$$1 \arcsec$ was selected for this study, it is worth noting that the outlined imaging approach can be applied to create images at other intermediate resolutions such as $1.5 \arcsec$, using the LoTSS survey data with international baselines. 

To assess the overall computational cost for imaging a $\sim 1 \arcsec$ resolution LoTSS-like survey, as a progression from the LoTSS survey with $6\arcsec$ resolution image products, several factors need to be taken into account. First, each pointing in the LoTSS survey is separated by approximately $2.6 \times 2.6$ $\mathrm{deg}^2$ \citep{2017A&A...598A.104S}, while our current image size is $2.5 \times 2.5$ $\mathrm{deg}^2$. This means that the computational cost for each pointing would be slightly higher than our current estimates. To make a $2.6 \times 2.6$ $\mathrm{deg}^2$ size image would necessitate imaging 23\,400 by 23\,400 pixels. The calibration costs would remain consistent, but final imaging would increase, resulting in a core-hour increase on the order of 100. Given that LoTSS has 3\,168 pointings to cover the entire northern sky, if a $1.2\arcsec$ resolution LoTSS-like survey shares the same pointing strategy, a total of approximately 165\,148\,000 core hours would be required. This cost is still in the same order as producing the LoTSS results.

If a uniform noise level across each pointing area is desired, we cannot use the same number of pointings as the LoTSS due to the small FOV of the international stations. Therefore, more pointings are needed to achieve a more uniform noise level and seamless imaging. An international HBA station has an FWHM of 2.59$^{\circ}$ at 120 MHz, 2.16$^{\circ}$ at 144 MHz and 1.85$^{\circ}$ at 168 MHz. Given that the separation between pointing centres for LoTSS was selected between FWHM/$\sqrt{2}$ and FWHM/1.2 \citep{2017A&A...598A.104S}, a separation between 1.31$^{\circ}$ and 1.54$^{\circ}$ for the highest frequency (168MHz) should be considered for the $1.2\arcsec$ resolution LoTSS-like survey. If we opt for 1.4$^{\circ}$ as the separation, approximately 12\,000 pointings would be needed. To achieve approximately uniform sensitivity across the entire sky during the mosaic procedure, we would use image sizes of $1.8 \times 1.8$ $\mathrm{deg}^2$, extending 30\% beyond the pointing separation. 

It's worth noting that the time resolution of the data used in this work was initially averaged from 1 second to 2 seconds. For survey-related imaging reaching or exceeding the FWHM of the international station's primary beam (2.15$^{\circ}$ at 144 MHz), we recommend using a 1-second time resolution to reduce the effects of time smearing, as the smearing effect could be larger, which would further increase the computational cost by an estimation of 10\%. The choice of a 1.4$^{\circ}$ field of view for a $\sim 1 \arcsec$ resolution LoTSS-like survey ensures that it remains within the FWHM of the international station's primary beam. Consequently, we can use the computational cost recorded in this work to estimate the computational requirements of such a survey without considering additional time-smearing effects.

To create a $1.8\times1.8$ $\mathrm{deg}^2$ size image at $1.2\arcsec$ would require imaging 16\,200 by 16\,200 pixels, reducing the computational cost to an estimated 45\,700 core hours. For 12\,000 pointings in total, the overall computational cost would be one order of magnitude higher than producing LoTSS results. This computational requirement is becoming increasingly feasible due to advances in software capabilities and the increasing power and availability of computing resources.

\section{Summary and Conclusions}

In this paper, we introduced the first wide ($6.45$ $\mathrm{deg}^2$) image at a resolution of $1.2\arcsec\times2\arcsec$ with a median noise of $\sim$80 $\mu \mathrm{Jy}$ ${\mathrm{ beam}}^{-1}$ using the International LOFAR High Band Antennas. This image was produced using an 8-hour observation at frequencies ranging from 120-168 MHz. We outlined our data reduction process, highlighting the most up-to-date ILT imaging steps used to produce the direction-dependent calibrated image. This resulted in the production of a radio source catalogue containing 2\,263 sources detected over the ELAIS-N1 field, using a peak intensity threshold of $5.5\sigma$. We have performed a cross-matching of our radio source catalogue with the LoTSS deep ELAIS-N1 field radio catalogue, resulting in the correction of flux density and positional inaccuracies. 

$\sim$80\% of the sources in the ELAIS-N1 Deep Fields catalogue are detected at $1.2\arcsec$ above $\sim$2 mJy, which is a factor of 2 larger than the number of sources detected at $0.3\arcsec$ in the Lockman Hole. This implies there is a wealth of information on $1.2\arcsec$ angular scales, and this catalogue represents a valuable resource for future studies of the ELAIS-N1 field. 

From a computational perspective, the production of one $\sim 1\arcsec$ resolution image from an 8-hour ILT observation takes approximately 52\,000 core hours, including multiple calibration and imaging steps. Notably, this represents only a fraction of the core hours required for sub-arcsecond imaging.

\begin{acknowledgements} 
    This research made use of \texttt{astropy}, a community-developed core Python package for astronomy (Astropy Collaboration 2013) hosted at \url{http://www.astropy.org/}. LOFAR designed and constructed by ASTRON has facilities in several countries, which are owned by various parties (each with their own funding sources), and are collectively operated by the International LOFAR Telescope (ILT) foundation under a joint scientific policy. The ILT resources have benefited from the following recent major funding sources: CNRS-INSU, Observatoire de Paris and Universite d’Orléans, France; BMBF, MIWF-NRW, MPG, Germany; Science Foundation Ireland (SFI), Department of Business, Enterprise and Innovation (DBEI), Ireland; NWO, the Netherlands; the Science and Technology Facilities Council, UK; and Ministry of Science and Higher Education, Poland; Istituto Nazionale di Astrofisica (INAF). This work made use of the Dutch national e-infrastructure with the support of the SURF Cooperative using grant no. EINF-1287. This publication is part of the project CORTEX (NWA.1160.18.316) of the research programme NWA-ORC which is (partly) financed by the Dutch Research Council (NWO). This project has received support from SURF and EGI-ACE. EGI-ACE receives funding from the European Union’s Horizon 2020 research and innovation programme under grant agreement No. 101017567. RJvW acknowledges support from the ERC Starting Grant ClusterWeb 804208. LKM is grateful for support from the Medical Research Council [MR/T042842/1]. PNB is grateful for support from the UK STFC via grant ST/V000594/1. FdG acknowledges the support of the ERC Consolidator Grant ULU 101086378. MB acknowledges support from INAF under the Large Grant 2022 funding scheme
(project "MeerKAT and LOFAR Team up: a Unique Radio Window on Galaxy/AGN co-Evolution. This publication is part of the project CORTEX (NWA.1160.18.316) of the research programme NWA-ORC which is (partly) financed by the Dutch Research Council (NWO). 
\end{acknowledgements}

\bibliographystyle{aa}
\bibliography{reference}
\begin{appendix}
\section{Images of selected extended sources}\label{appendix}

\begin{figure*}[h!]
\includegraphics[width=\textwidth]{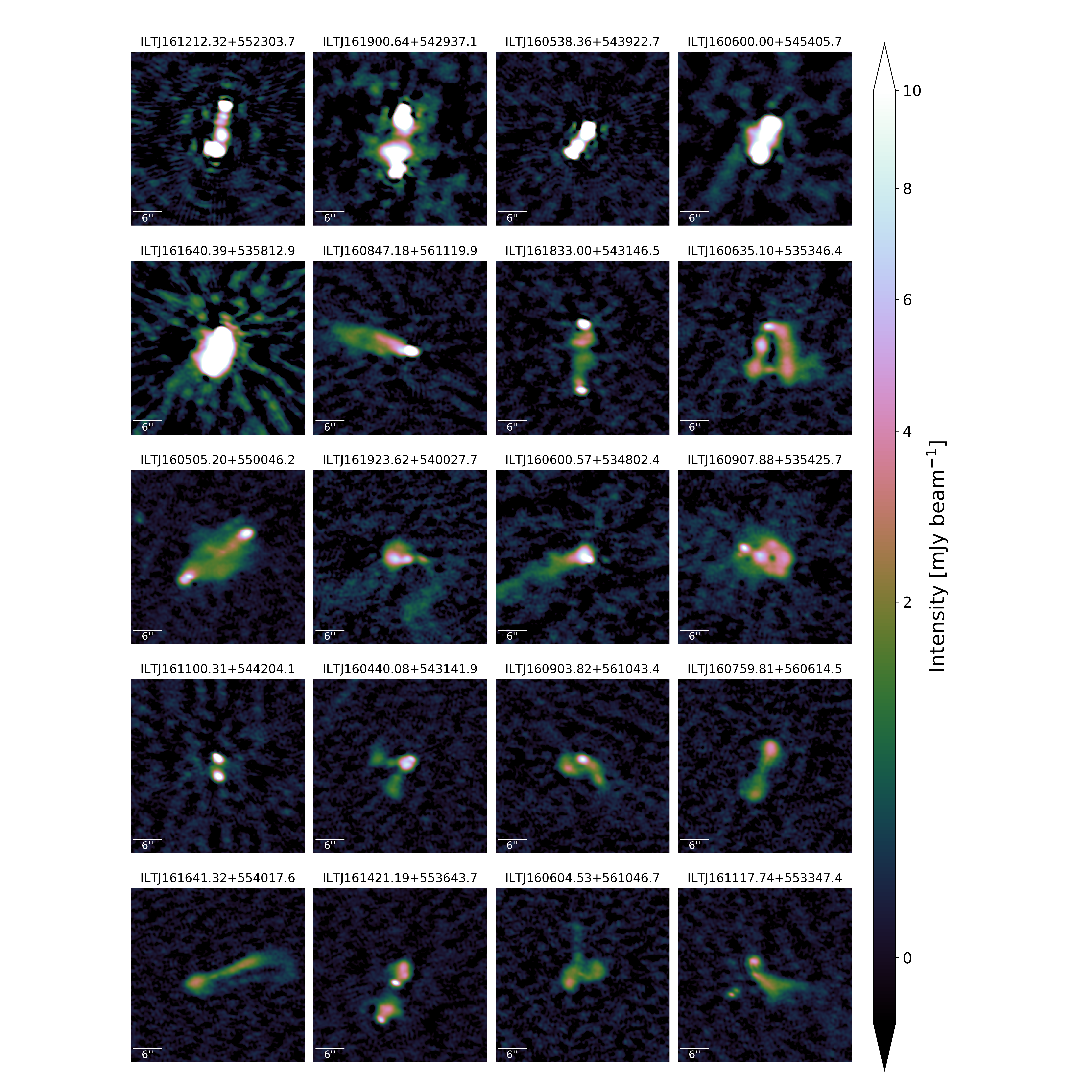}
\captionof{figure}{Selected 20 extended sources from the $1.2\arcsec$ resolution radio catalogue with a peak flux greater than 2 mJy ${\mathrm{beam}}^{-1}$ and a FWHM of their major axis larger than 7.2 arcseconds.}\label{Fig:cutout_extended_image}%
\end{figure*}

\begin{figure*}[h!]
\includegraphics[width=\textwidth]{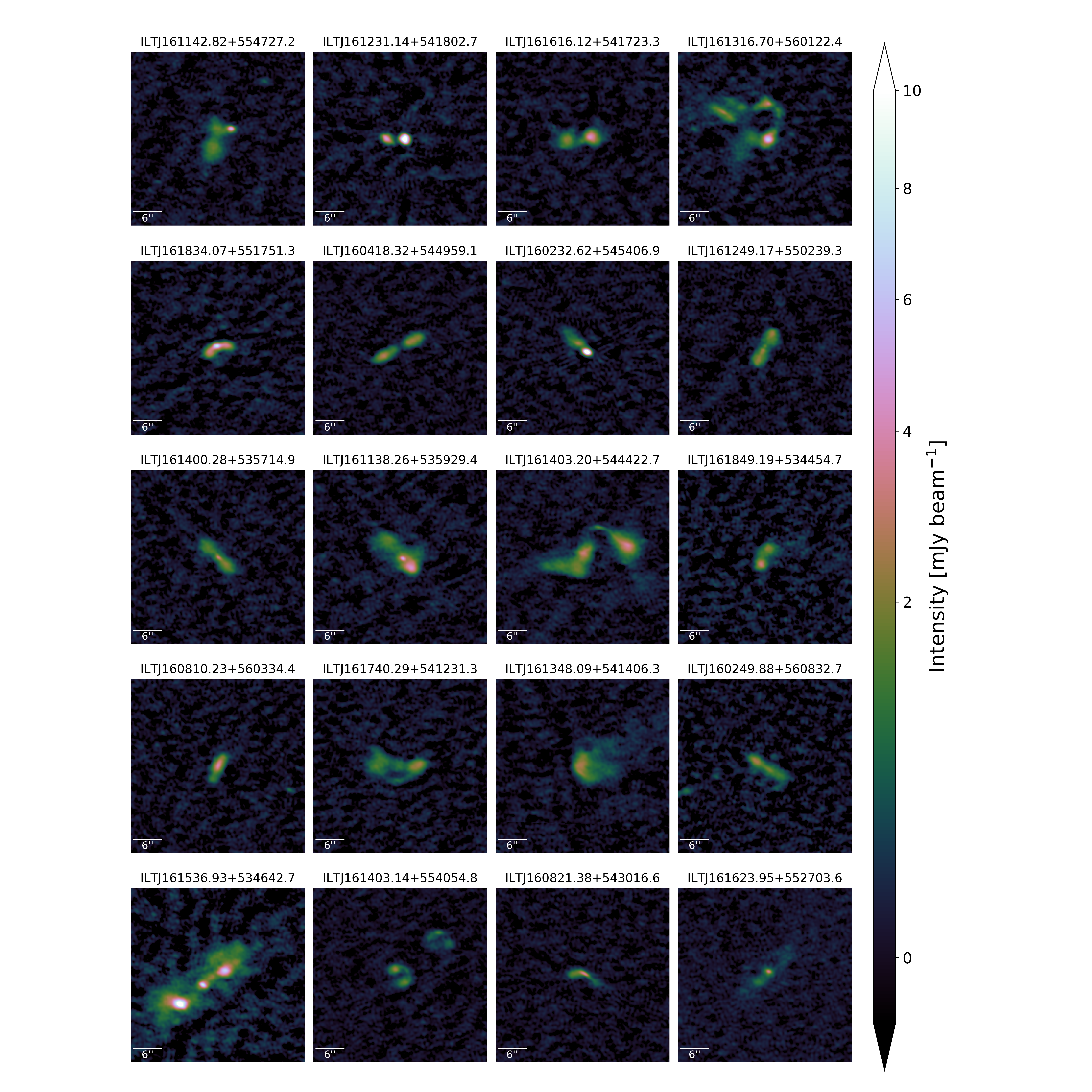}
\captionof{figure}{Selected another 20 extended sources from the $1.2\arcsec$ radio catalogue with a peak flux greater than 2 mJy ${\mathrm{beam}}^{-1}$ and a FWHM of their major axis larger than 7.2 arcseconds.}\label{Fig:cutout_extended_image1}%
\end{figure*}

Cut-out images of 40 selected extended sources from our $1.2\arcsec$ resolution radio catalogue are displayed in Figures \ref{Fig:cutout_extended_image} and \ref{Fig:cutout_extended_image1}. All 40 sources have peak fluxes greater than 2 mJy ${\mathrm{beam}}^{-1}$ and FWHMs of their major axes larger than 7.2 arcseconds. Each image is extracted from our final image after astrometric and flux-scale correction, with dimensions of 72 by 72 arcseconds.

\section{Facet-dependent analysis for astrometric and flux density correction}\label{appendix:2}

\begin{figure*}
   \centering
   \includegraphics[width=0.6\textwidth]{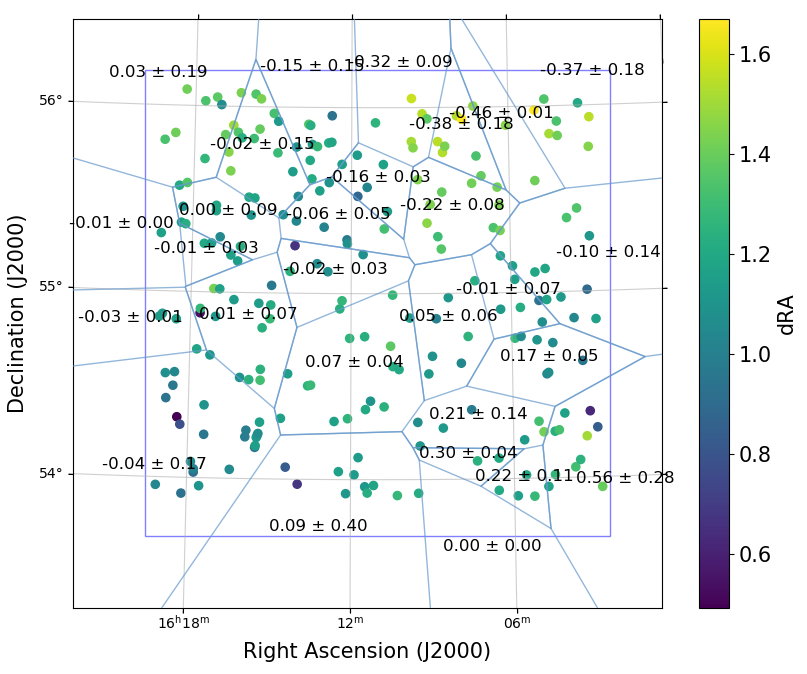}
   \caption{The 231 sources selected for astrometric correction are plotted as coloured dots on the image plane, with each imaging facet marked by blue polygons. The mean of the dRA values for all the sources within a given facet, along with the standard deviation, are displayed.} \label{Fig:position_polygon}%
\end{figure*}

\subsection{Astrometric correction}

\begin{figure*}
   \centering
   \includegraphics[width=0.6\textwidth]{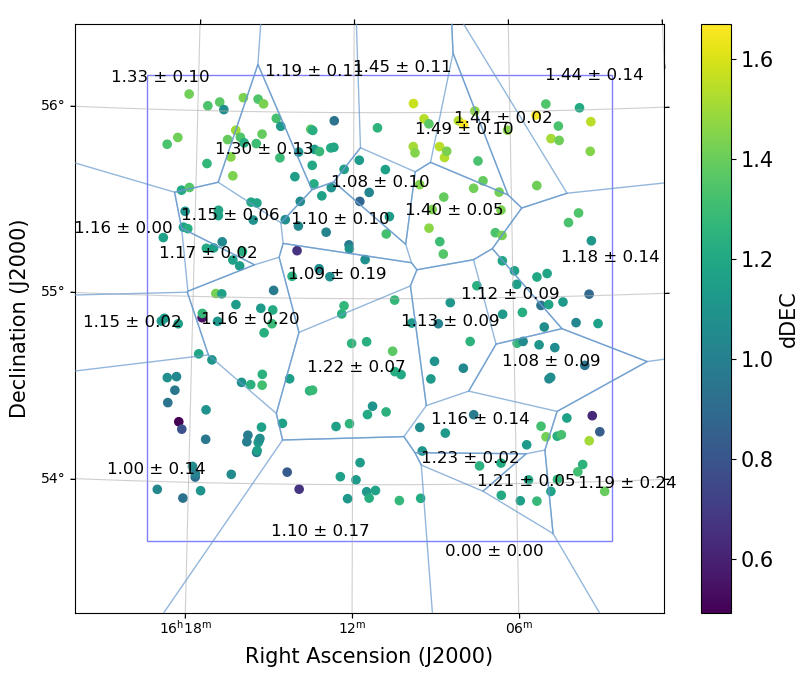}
   \caption{The 231 sources selected for astrometric correction are plotted as coloured dots on the image plane, with each imaging facet marked by blue polygons. The average of the dDEC values and their standard deviations are marked within each facet. If we subtract the astrometric offset derived for DEC in this work (1.9134) from the mean values of each facet, the mean values of each facet are well below $1.2\arcsec$, the resolution of our image.} \label{Fig:position_polygon2}%
\end{figure*}

To investigate facet-dependent variations in positional errors, we plotted the positions of 231 selected sources used for astrometric correction in Section \ref{subsec: astrometry} on the image plane, shown in Figure \ref{Fig:position_polygon} and \ref{Fig:position_polygon2}. Each imaging facet is delineated by blue polygons, with coloured dots indicating dRA (Figure \ref{Fig:position_polygon}) and dDEC (Figure \ref{Fig:position_polygon2}) values for the sources. For each facet, we computed the mean and standard deviation of the dRA and dDEC values for all selected sources within that facet. These values were then marked inside the corresponding facet on the plot. The facet marked by `NA' indicates that there are no samples selected from this facet, hence no average of the dRA or dDEC can be calculated for that facet.

In Figure \ref{Fig:position_polygon} and \ref{Fig:position_polygon2}, we can see that there is variation of the dRA and dDEC from facet to facet due to the facet-based imaging approach. However, the small variation across the field is well below the resolution of our image.

\begin{figure*}
   \centering
   \includegraphics[width=0.9\textwidth]{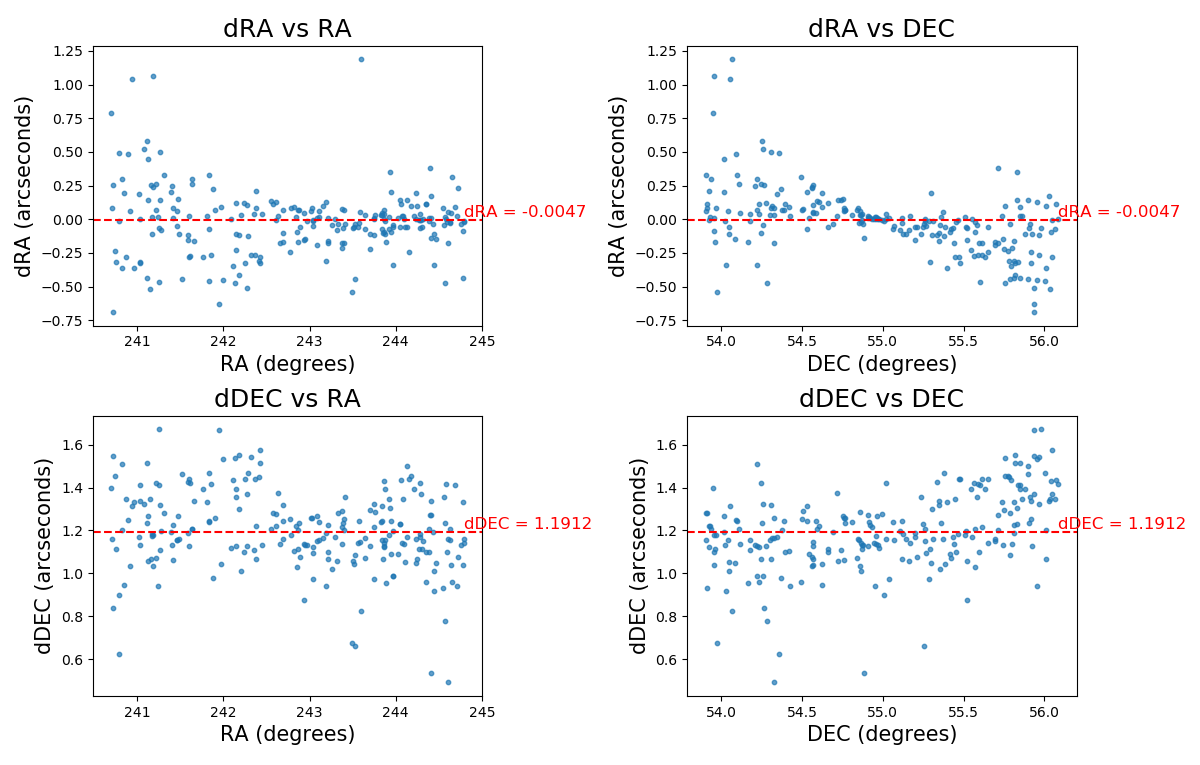}›
   \caption{The positional errors dRA and dDEC of the selected 231 sources are plotted against their positions RA and DEC in the four subplots.} \label{Fig:dDEC_dRA_RA_DEC}%
\end{figure*}

We also plotted the positional errors dRA and dDEC of the same 231 sources against their positions RA and DEC in the four subplots in Figure \ref{Fig:dDEC_dRA_RA_DEC}. The standard deviation of the positional errors tends to be slightly lower at the image centre compared to the image edge. While the trend in the astrometric offsets appears significant, they remain well below the resolution of our image. Therefore, we decided to apply only this one astrometric correction to RA and DEC.

\subsection{Flux density correction}

\begin{figure*}
   \centering
   \includegraphics[width=0.7\textwidth]{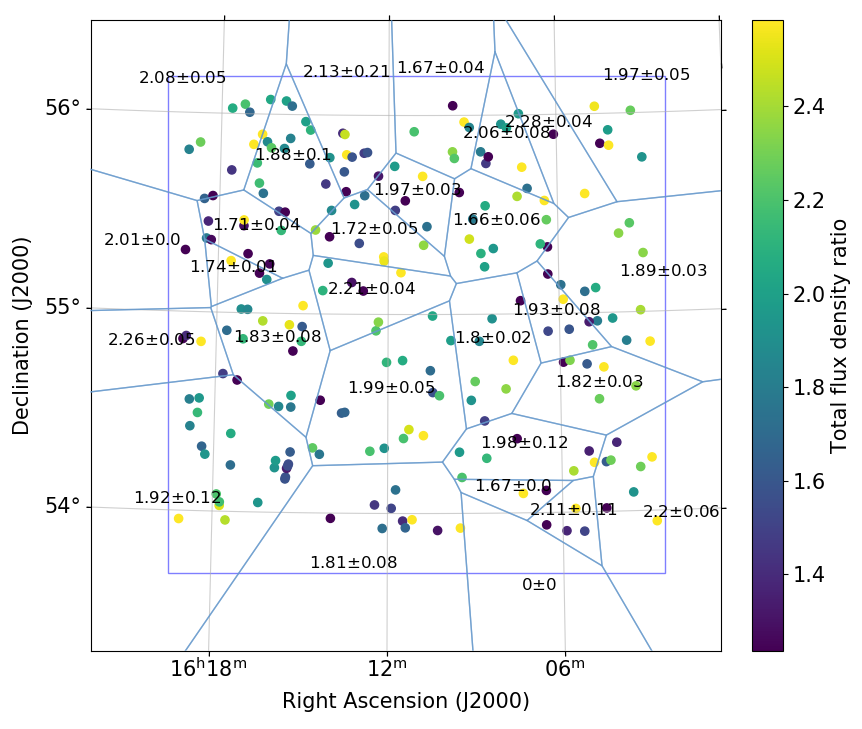}›
   \caption{Sources selected to determine the flux density scaling factor in Section \ref{subsec:fluxscale} are plotted as coloured dots on the image plane, with each imaging facet marked by blue polygons. The average values of the flux density ratios for all the sources within a given facet is displayed inside the facet along with the standard deviations.} \label{Fig:flux_scale_polygon}%
\end{figure*}

To investigate potential systematic facet-dependent effects on the flux scaling factor, we plotted the 223 sources selected to validate the flux density scaling factor in Section \ref{subsec:fluxscale} as coloured dots on the image plane. In Figure \ref{Fig:flux_scale_polygon}, each imaging facet is marked by blue polygons. The dot colours represent the flux density ratio, calculated as the ratio between our $1.2\arcsec$ resolution radio catalogue and the $6\arcsec$ resolution deep field radio catalogue. For each facet, we computed the flux density ratios for all selected sources within that facet, calculated their mean and standard deviation, and then marked these values inside the facet. The facet marked by `NA' indicates that there are no samples selected from this facet, hence no average of the ratios can be calculated for that facet. Upon comparing the average values across all facets, we did not observe systematic facet-dependent effects.

\begin{figure*}
   \centering
   \includegraphics[width=\textwidth]{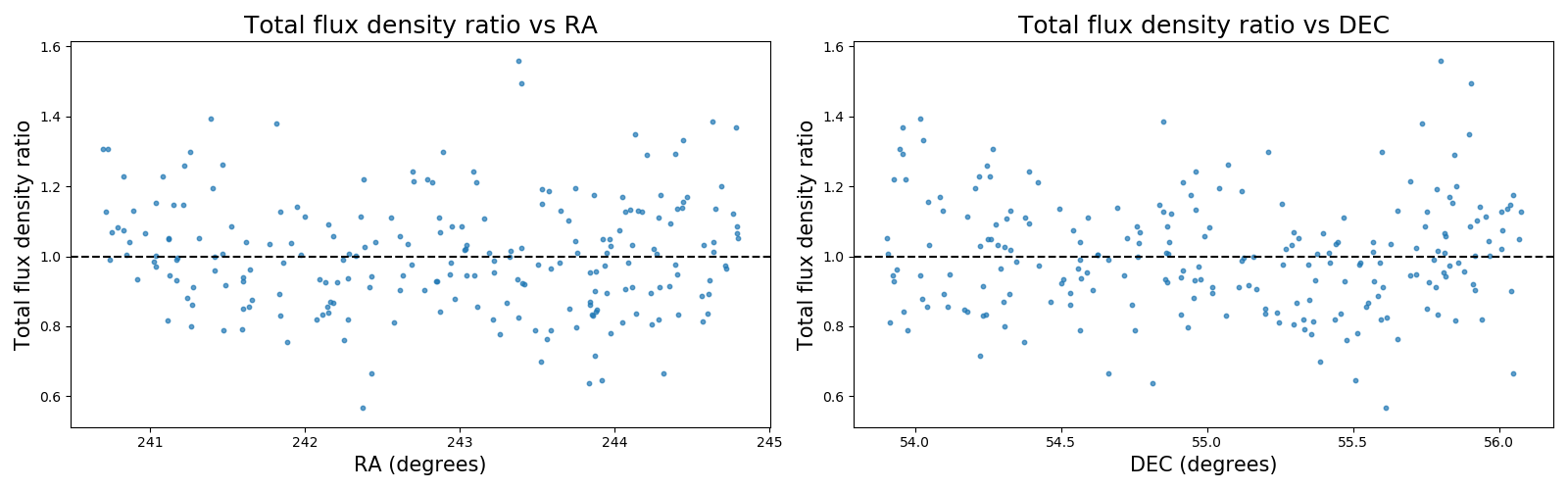}›
   \caption{The flux density ratio of the selected 223 sources are plotted against their positions RA and DEC in the two plots.} \label{Fig:ratio_RA_DEC}%
\end{figure*}

Additionally, we plotted the total flux density ratio versus RA and DEC. As shown in Figure \ref{Fig:ratio_RA_DEC}, we did not observe any obvious trends in these two plots.

\end{appendix}

\end{document}